\documentclass[times,astrobib,amssymb,usenatbib]{mn2e}
\usepackage{graphicx, subfigure}
\usepackage{graphicx}
\usepackage{txfonts}
\usepackage{natbib}
\pdfoutput=1 
\usepackage{psfig}
\usepackage{epsf}
\begin{document}

\title[YSOs and filamentary structures in the inner Galaxy]
{Study of young stellar objects and associated filamentary structures in the inner Galaxy}
\author[B Bhavya et al.]{B Bhavya$^{1}$\thanks{E-mail:bhavyab@cusat.ac.in}, 
Annapurni Subramaniam$^{2}$\thanks{E-mail:purni@iiap.res.in}, V C Kuriakose$^{1}$
\\
$^{1}$Department of Physics,Cochin University of Science \& Technology, Kochi 682\,022, India \\
$^{2}$Indian Institute of Astrophysics, Bangalore 560034, India \\}

\maketitle

\begin{abstract}
Young Stellar Objects (YSOs) in the inner Galactic region $10^0 < l < 15^0$ and 
$-1^0 < b < 1^0$ are studied using GLIMPSE images and GLIMPSE data 
catalogue. A total number of 1107 Class I 
and 1566 Class II sources are identified in this Galactic region. 
With the help of GLIMPSE 5.8 $\mu$m \& 8 $\mu$m images, 
we have identified the presence of 10 major star 
forming sites in the Galactic midplane, of which 8 of them are filamentary 
while 2 are possible clusters 
of Class I $\&$ II sources. The length of the identified filaments are 
estimated as 8'-33' ($\sim$ 9 - 56 pc). Occurrence of Hub-Filamentary System (HFS) 
is observed in many filamentary 
star forming sites. Most of the Class I sources are found to be aligned 
along the length of these filamentary structures, while Class II sources have 
a random distribution. 
Mass and age distribution of 425 Class I and 241 Class II sources associated with 
filaments \& clusters 
are studied through their SED analysis. Most of the Class I sources detected 
have mass $>$ 8M$_\odot$, while Class II sources have relatively low mass
regime. Class I sources have ages $\le$ 0.5 Myr, while Class II sources 
have ages in the range $\sim$0.1 - 3 Myr. Along with the help of 
high mass star forming tracers, we 
demonstrate that the 10 regions studied here are forming a large 
number of high-mass stars. 
\end{abstract}


\section{Introduction}
Star formation processes in the inner Galactic regions are important to 
understand as it happens in an environment of relatively 
higher density and metallicity. 
Apart from that these are sites where high mass star
 formation happens vigorously. It is interesting to see that the galactic 
region from $10^0 < l < 15^0$ nurtures many features like filamentary 
structures, bubbles, HII regions which are observed only at longer 
wavelengths indicating that the region is highly dynamic, active and 
young in nature. Heavy obscuration of visible light blocks the study of 
inner galaxy using optical wavelengths. Owing to its high sensitivity 
and angular resolution, {\it Spitzer} 
(\citet{wer04}) revolutionized our view of inner galaxy and enabled us to
detect features which were not previously identified using optical 
wavelengths. The results are far reliable and accurate compared to previous 
poor resolution far-IR and millimeter observations. 
Mid Infra-Red (MIR) observations, which are less affected by dust extinction 
compared to near-IR and optical wavelengths, done with Infra
Red Array Camera (IRAC) of Spitzer (\citet{fazio04}) lead a deeper 
understanding of early star formation scenario. The Galactic midplane 
which comprises the major sites 
of massive star formation is 
mapped with Science Legacy Program GLIMPSE surveys. GLIMPSE I 
(\citet{benj03}) survey 
imaged the inner Galaxy from longitudes $l$=10$^0$ to 65$^0$ and latitude 
$|b|\le 1^0$ with resolution $<$ 2$"$ in all IRAC bands. 

Relative to the low-mass star formation processes, the high mass star 
formation mechanism is poorly understood. Some of the reasons for the paucity of
such studies are: high 
mass stars have a short pre-main sequence 
time scale, most of the high-mass star forming regions are distant, 
difficulty in observing 
earliest stages of high-mass star forming processes, difficulty in 
getting a bonafide sample of 
high-mass stars, highly embedded nature of massive star forming sites make them 
difficult to resolve and locate. Hence, it has been difficult to 
characterize high-mass ($\ge$ 8M$_\odot$) star forming processes. 
Understanding high mass protostars are highly essential in 
the study of universal initial mass function. 
In this paper, we present a study of star forming regions 
at low Galactic latitudes. This region is
suggested to be undergoing high mass star formation as revealed 
by previous studies. 
The region of our interest is the inner Galactic plane, 
in the longitude-latitude range, $10^0 < l < 15^0$ and $-1^0 < b < 1^0$, 
and close to M17, W33 and W31 star forming regions. 
This region is located to the south-west of M17 
which harbors many features which 
are not detectable in optical, but shows prominent features in longer 
wavelengths including mid-IR. CO map given by 
\citet{san86} shows the presence of molecular clouds in this region.
Several regions in this longitude-latitude range have been previously 
studied by others. Many of the studies are based on the HII region complexes 
such as, W31 (G10.2-0.3 \& G10.3-0.1) (\citet{wil74}, \citet{caswell75},  
\citet{corbel04}, \citet{kimkoo02}, \citet{furness10}, \citet{beu11}) 
and W33 (\citet{goss78}, \citet{soifer79}, \citet{stier82}, \citet{gold83}, 
\citet{haschisk83}, \citet{messi11}) and also other ultra compact HII regions, 
G10.6-0.4 (\citet{persson12}, \citet{gerin10}, \citet{neufeld10}, 
\citet{klassen11}, \citet{liu11}, 
\citet{sollins05}), G10.47+0.03 (\citet{rolffs11}, \citet{pascucci04}, \citet{olmi96}), 
G10.62-0.38 (\citet{beltran11}) 
etc.
High mass star formation in the infra-red dark cloud G11.11-0.12 (Region 9 
in our study), which is at a kinematic distance of 3.6 kpc (\citet{clemens86}), 
has 
been studied using water and methanol masers by \citet{pillai06}, 
using Herschel data by \citet{henning10} and more recently by 
\citet{gomez11}. 
Though all the above regions are located in the first galactic quadrant 
covering the near 
3-Kpc arm, studies do not discuss their connection with the 3-Kpc arm. 
According to \citet{downes80} W31 complex is a part of 3-Kpc arm, 
while \citet{dame08} find that the velocity peaks of W31 are not in 
accordance with the general 3-Kpc arm kinematics.
The detection of ultra compact HII regions implies the primitive 
nature of this region. This might suggest that 
a large amount of molecular cloud in the inner galaxy 
is triggered to form stars, similar to ''global triggering'',  
as indicated by \citet{povich10} in the case of M17SWex. 
This region is a part  
of several surveys like CO maps, methanol masers, infra red dark clouds
(IRDCs), star-less clumps, HII regions etc. We have used the results from 
some of the previous surveys as tracers of high mass star formation and 
to resolve the uncertainty in distance towards these regions. 

Elongated structures of parsec scales seen in star forming complexes 
have been extensively studied by \citet{myers09}, who analyzed the  
general properties of elongated structures observed 
in deep optical, near-IR, CO mapping, IRAC and GLIMPSE/MIPSGAL images. 
According to which, a "hub" is termed as a central blob of 
high column density, with peak column density 10$^{22-23}$ cm$^{-2}$ 
and filament denotes associated feature of low-column density. Smaller hubs 
tend to be relatively round, with fewer stars, lower column density and with a 
few radiating filaments. 
Larger hubs are more elongated, with more stars, 
higher column density and with 5-10 filaments. These filaments are nearly 
parallel to each other, directed along the short axis of the hub, with similar 
spacing and direction forming a "hub-filamentary system" (HFS). 
HFS is seen in dust emission and absorption, and in molecular
line emission. It is seen in optical dark clouds within a few
hundred pc, and in IRDCs at distances $\sim$3 kpc.
In this region, we notice filamentary features, with a few showing HFS. 

Here we make use of the point-source catalog and image cut-outs
from GLIMPSE I survey to 
demonstrate the young stellar content and features in the inner 
galactic mid-plane. The YSO spectral energy 
distribution deviates at mid-IR wavelengths from normal 
photospheric emission.
The Spitzer-IRAC sensitivity to mid-IR emission 
makes it the best tool to identify and characterize YSOs. Sources 
with excess mid-IR emission are classified as Class I (still 
embedded and accreting from dense 
spherical envelopes) and Class II (slightly more evolved pre-main 
sequence stars with circumstellar disks) 
(\citet{lada06}, \citet{guter08}). The mass and the age 
range of YSOs in each region
are estimated and used to understand the ongoing star formation 
and the filamentary structure. \\

This paper is structured as follows. Section 2 describes the archival 
data catalogues we 
used for the study of galactic central regions. In Section 3 we detail 
the analysis and results, 
which includes Classification of IRAC sources, Spectral Energy Distribution 
fitting for the YSOs, Identification of Filamentary 
structures \& Clustering with its descriptions, 
Spatial distribution of YSOs and tracers of high-mass 
star formation. In Section 4
we include a brief discussion and summarize our main findings in section 5. 

\section{Archival Data catalogues used for this study}
In order to understand the details of star formation in the 
region of interest, we combine the following archival data with the YSOs 
identified in this study. As these sources are
tracers of high mass star formation, these will be used to compare 
the masses and ages of the YSOs studied here.
Since we do not have any distance estimate towards the YSOs studied here, we use 
the distance estimate for the sources
in the archival data to obtain a possible range of distance to the YSOs.
As some of the sources are mentioned in the literature to be associated 
with the inner spiral arms of the Galaxy, we combine this 
information to connect the features identified in this study with 
the spiral structure of the inner Galaxy. \\
{\it \bf Infra Red Dark Clouds (IRDCs): } \\
IRDCs, which appear as silhouettes in the general galactic mid-IR background 
are the precursors of cluster forming molecular clumps (\citet{carey98}, 
\citet{jack08}) 
and IRDC cores host the very earliest phases of high mass star formation 
(\citet{rath05}, \citet{rath06}, \citet{rath07}, \citet{simon06}). 
We have taken the catalogue 
from  
\citet{peretto09} which provides a complete sample of IRDCs 
in the galactic range $10^0 < |l| < 65^0$ and $|b| < 1^0$
using GLIMPSE and MIPSGAL survey. \\
{\it \bf Methanol Masers:}\\
According to \citet{green09}, the Methanol maser transition at 6.7 GHz 
have been observed towards 
early hot core phases of star formation processes and found to be associated 
with tracers of high mass star formation in IRDCs (\citet{ellingsen06}) and 
Extended Green Objects (EGOs) (\citet{cygano08}). 
Using Methanol Multibeam (MMB) Survey, 
\citet{green09} and \citet{green10} conducted a search for methanol 
masers and showed that significant star formation is 
happening 
in the 3-Kpc arm, which is existing within the $15^0$ of Galactic center. 
MMB detected more than 200 methanol 
masers in the region $15^0 < l < -15^0$ and 49 of them have velocity peak
matching that 
of near and far 3-Kpc arms. Among them 52 sources are located in our 
region of interest, of which, 4 sources are identified as 3-Kpc arm 
signatures and 2 sources as far arm features by \citet{green10}.\\
{\it \bf Star-less clumps:}\\
Using ATLASGAL survey at 870 $\mu$m, \citet{tacken12}, made a search for 
dense gas condensations. Along with GLIMPSE catalogue and 24$\mu$m MIPSGAL 
images, this survey showed the existence of starless cores which may 
form high mass stars, 
in the galactic region $10^0 < l < 20^0$ and $|b| < 1^0$. The catalogue also 
provides distances (both far and near) towards these objects.\\
{\it \bf IR bubbles/HII regions:} \\ 
IR bubbles, as being the most spectacular objects in the GLIMPSE/MIPSGAL 
images, 
are the HII regions produced by radiation and wind 
from O and early B type stars of age $\sim$$10^6$ yr (\citet{church09}) and 
therefore located at sites of recent massive star formation.
\citet{and11} presents the catalogue of HII regions in the
galactic region $343^0 \le l \le 67^0$  and $|b| \le 1^0$. \\
{\it \bf Radio sources:} \\
Galactic radio sources are taken from radio continuum survey (CORNISH survey)  
by \citet{purce13}, targeting UCHIIs in high mass star formation. 
CORNISH project covers northern GLIMPSE region 
($10^0 < l < 65^0$) using the Very Large Array at 5 GHz. 
We identified 10 such radio sources to be located in our $l-b$ range.

We also show the location of 2 high mass protostellar objects (HMPOs) 
from \citet{grave09} and Extended Green Objects (EGOs) as tracers of 
massive young stellar objects 
(MYSOs) from \citet{cygano08}, (4 'likely' MYSO ouflows and 1 'possible' MYSO 
outflow), situated in our region of study. 
The spatial association of these sources (mainly as tracers of high mass star 
formation) and YSOs identified in our analysis will be 
discussed in later sections.

\section{Analysis \& Results}
\subsection{Classification of IRAC sources}
More than 5 million sources from GLIMPSE I survey in the 
longitude-latitude range $l = 10^0 - 15^0$; $-1^0 < b < 1^0$
are examined to identify candidate YSOs. 
As only Spitzer-IRAC magnitudes are used in the identification 
of YSOs, in order to improve the reliability, 
we have taken only those IRAC 
sources with error less than 0.1 mag in all IRAC bands.
The contamination from non-YSO sources and 
the effect of reddening are to be eliminated to create a bonafide sample of 
YSOs. We follow the 
\citet{guter09} (hereafter G09) procedure to remove background contamination and 
to select Class I and Class II sources. Brief outline of 5 steps in identifying 
YSOs are described here. 

G09 proposed an empirical scheme for identifying and
classifying YSOs on the 
basis of mid-IR magnitudes and colors eliminating possible contaminants. 
According to \citet{stern05}, galaxies dominated by PAH emission 
(normal star-forming galaxies) show excess in 4.5 $\mu$m and 5.8 $\mu$m 
band passes and broad line AGNs, having non stellar spectral energy 
distributions, 
exhibit MIR color
indices similar to YSOs. Only very few (3) PAH emission sources and 
no AGN sources are 
identified in our sample by giving respective color cuts based on G09. 

The high velocity outflows from protostars colliding with surrounding 
molecular cloud cause unresolved blobs of shock emission and are sensitive 
to the 4.5 $\mu$m IRAC band as it covers molecular hydrogen emission lines. 
By giving G09 constraints in the [4.5]-[5.8] Vs [3.6]-[4.5] 
color space, we detect 19 sources as shock emission sources and are removed. 
Additional contaminants are spurious excess
emission 
in the 5.8 $\mu$m and 8.0 $\mu$m band passes 
caused by PAH emission sources, which contaminate the photometric apertures of 
some faint field stars. There were 13563 of such contaminants. 

After removing the above mentioned contaminants, sources which satisfy 
the conditions of 
protostellar colors are classified as Class I sources. 
Figure 1{\sl(Left)} shows the Class 1 sources (red triangles), other 
contaminants such as shock emission sources (blue) and PAH aperture 
contaminants (green) and the unclassified remaining sources (black). 
First these Class I sources are extracted. Later, from the remaining sources, 
those with 
consistent colors of pre-main sequence stars having circumstellar disk 
are classified as Class II sources.
Figure 1{\sl(Right)} 
shows Class II sources (red) and unclassified sources (black) which mostly 
contain stars close to main sequence. There are 1107 Class I sources 
and 1566 Class II sources identified in the region of interest. 
GLIMPSE catalogue also provides 2MASS magnitudes for these 
sources. All these sources have 2MASS magnitudes, but only 
103 Class I and 1090 Class II sources have 2MASS magnitudes 
with error less than 0.1 mag. We have not de-reddened IRAC colors. 
We caution that reddening can affect the 
the number of Class I\&II sources identified. The extinction vectors 
(\citet{flaherty07}) 
corresponding to A$_k$=5 mag are shown in the Figure 1. 
As our study does not aim to obtain a complete 
sensus of YSOs in this region, but understand the 
nature and location of early star formation, we do not attempt to achieve the 
completeness limit in detecting YSOs. Since an upper limit for the error 
in the IRAC magnitudes are set, our estimation of YSOs is a lower limit 
of the number of YSOs present in this location. Sources identified 
here are likely
to be genuine detections and would help us to derive reliable properties 
of these sources as well the star formation sites.

We point out the caveats and drawbacks of this study. 
Since only those sources with sufficient flux in all IRAC bands are 
considered in this study,
 faint low-mass ClassI\&II sources, 
evolved Class II sources and distantly located sources will be missed. 
Reduced number of these sources
in our study is due to the sample selection and not due to the absence of 
these sources in the region. 
There will also be source contamination from edge-on disk sources 
and reddened Class II sources which would mimic their colors as 
Class I sources. Gutermuth et al. (2009) gives an upper limit 
for the edge-on disk source confusion as 3.3$\pm$1.5$\%$. \citet{kryukova12} 
showed that there will be 4$\%$ contamination due to 
edge-on sources in their YSO sample. 
We also assume a similar fraction of edge-on sources in 
our sample. The major impact of not using MIPSGAL data is that 
we are unable to identify reddened Class II sources.

In order to
understand the distribution of these sources,
the location of Class I and Class II sources in the region studied are 
shown in Figure 2. Location of M17, W31 and W33 are also shown. 
It can be seen that the Class I sources are found to be located as groupings, 
whereas the Class II sources are located all over the region. 
The class I sources are distributed very close to the galactic plane, 
except in the M17SWex region. Specific locations of Class I groups 
can be identified and found to coincide with the location of other 
tracers of high-mass star formation. 
The Class II sources do not show any preferential confinement to the 
galactic plane. They are distributed randomly in the region, except for a 
couple of clusters. Thus, the Class I sources are likely to be associated with 
the sites of high-mass star formation, most of which are located along the galactic 
plane in the $l$-$b$ range studied. 

\subsection{Filamentary structures and Hub-Filamentary System (HFS)}
The 5.8 $\mu$m \& 8 $\mu$m images from GLIMPSE reveal the presence of 
filamentary structures in the region, which appear as dark 
patches in mid and far IR images of the region.
Close inspection of the images reveal several tiny filaments 
emanating from the main filament forming a
hub-filamentary system (HFS). Class I sources are found to be aligned as a string 
along the length of these filamentary structures. We identify 8 such  
filamentary structures of star forming regions and 
2 candidate clusters of Class I $\&$ Class II sources 
in the entire region, each of which will be described 
in detail in section 3.5. 
Filamentary structures are assumed to be primitive star forming sites; 
whereas the candidate clusters are slightly evolved.
The location of these identified regions are marked in Figure 2.
We have defined boxes to differentiate the 
filamentary regions \& clusters from background. 
Though the sizes of the boxes do not limit the actual 
physical extent of each region, we have 
considered only those sources inside the boxes for further analysis. 
Size and shape of the boxes which define the boundary of the regions are defined
to include most part of the filamentary structure as seen in the images.  
We study the YSOs in these structures and correlate the location of Class 
I and II YSOs with the filamentary structures to get the 
nature of star formation they host. 
Though the Class II sources are randomly distributed in these regions 
(except for the regions 
near W33 and near M17SWex where we can see a preferential 
clustering of Class II sources), 
Class I sources are either clumpy or closely 
associated with the filaments. In the entire region, 
we detect more number of Class II sources, 
compared to the Class I sources, with a ratio of 0.7.

In all the filamentary regions, there is a higher 
concentration of Class I sources along the filaments. 
This can be due to either the 
reddened Class II sources appearing as Class I sources or intrinsically, 
Class I sources are more there. 
These filaments of dark clouds can be the precursors of massive star formation
and progenitors of young clusters as noted by previous studies. 
The catalogue of IRDCs taken from \citet{peretto09} in the same $l$-$b$ range
is used to see whether the IRDCs and YSOs are co-located.
When positional match is done between YSOs and IRDCs, their central 
coordinates do not match within 2" separation. 
But we see the association of YSOs and IRDCs from 3" separation onwards. 
In order to quantify the association of YSOs with the known IRDCs, we computed the 
distance of Class I/II sources from the IRDCs. The histogram (Figure 3) 
shows the summary of this
 estimate for all the IRDCs. Up to a distance of 90" from IRDCs, 
the number of Class I sources are more 
than the number of Class II sources. Beyond this, the number of Class I and 
Class II sources are similar. \citet{peretto09} gives the physical 
extent of each of the IRDCs, which comes 
greater than 5" (in the range 30"-50"; in some cases more than ~200"). Hence,
it can be assumed that at least some of the YSOs are associated 
with the same region where the IRDCs are also found. In general, 
wherever we find a high 
density of Class I sources, number of IRDCs are also found to be more.
This once again proves that the filamentary regions identified here are 
probable sites of high mass star formation.
\subsection{Distances}
The region in southwest of M17, $9^0 <l<14^0$, $-0.2^0<b<-0.45^0$ is termed as
3-Kpc 
arm (\citet{rougoor60}, \citet{clemens86}) which contains several molecular 
clouds; where the near 3-Kpc arm is at a distance 
of 5.2 kpc from Sun. Most of the star formation studies in the 3-Kpc 
arm are focused on the 
other end of the arm which extends in the $l$ range $340^0-360^0$ 
(4th Galactic quadrant). 
The association of the filamentary structures in our study 
to near 3-Kpc arm has been 
checked by comparing other studies in this region.
 In order to physically 
associate sources they have to be coherent in {\it lbv} space. 
The latitude range $10^0 < l < 15^0$ meets several spiral arms as crossing over
in the 
galaxy, the velocity measurements in this $l$ range will give ambiguous results. 
Though our region of study is an extent of M17SWex in $l$-$b$ space; 
it is more close to galactic plane ($b$ is less compared to M17 and M17Swex). 
So we assume that this region is not physically associated with 
Sagittarius arm, but might be a part of other inner arms. 
M17 and M17SWex are at a distance of 2.1 kpc in the Sagittarius arm (\citet{povich10}). 
The previous distance estimations towards W31 and W33 star forming 
regions give a range of values (3.3 -  7 kpc 
towards W31 and 2.4 -  7 kpc towards W33).
Keeping aside the distance ambiguity, 
the study of this region is of importance as it can give insights into 
initial states and characteristics of 
the formation of massive stars, OB associations and stellar clusters. 

Since the uncertainty related to the distance exists 
for the identified sources in the entire region, while fitting SED, 
we have given a common distance range of 4-6 kpc in the input of 
SED fitting tool 
(except for regions 1 $\&$ 2 and for region 9 as 2.1 kpc (\citet{povich10}) 
and 3.6 kpc (\citet{henning10}) respectively, with a $\sim$10$\%$ 
uncertainty in range (Table 1.). 
\subsection{Spectral Energy Distribution}
In order to characterize the nature of YSOs, we construct spectral energy 
distributions (SEDs) for 
Class I and Class II objects identified in the filamentary regions and clusters. 
We follow the online YSO SED fitting tool developed 
by \citet{robita07} to estimate the physical properties 
such as mass, age, accretion rate of disks and envelopes of YSOs.  
In the Robitaille et al. (2007) models, 
masses are sampled between 0.1 to 50M$_\odot$ and ages between 10$^3$ to 10$^7$ Myr. 
For each set of mass and age, temperature and 
radius are found by interpolating pre-main sequence evolutionary tracks of 
\citet{seiss00} and \citet{bernas96}. 
Once the stellar parameters are determined, 
values of disk and envelope parameters are sampled from ranges 
that are functions of evolutionary age of the central source, as well as 
functions of stellar masses in certain cases. Parameters corresponding 
to the model which fits the observed flux values with $\chi^2_{min}$ 
are taken as the YSO parameters.
Parameters of models which satisfy the criteria 
$\chi^2_{min}$ - $\chi^2_{best}$ $<$ 3 
where $\chi^2$ is the statistical
goodness of fit parameter per data point, are used to estimate the 
error in the estimated parameters. 

SED analysis is carried out only for 425 Class I and 241 Class II 
sources which are detected within boundary of the 10 identified regions. 
We have assumed an A$_v$ range of 1-40 mag for these sources. 
SEDs are constructed using 2MASS and IRAC magnitudes 
(wavelength range 0.1-8 $\mu$m) for some sources, 
whereas only IRAC magnitudes are considered for other sources. 
This is because, some YSOs have error more than 0.1 mag in the 2MASS magnitudes, 
which give unrealistic SED fittings. 
This means that for each region a few SEDs
are made from  IRAC and 2MASS magnitudes, whereas the 
remaining SEDs are constructed with IRAC magnitudes only.
Figure 4 shows examples of SEDs of two Class I and two Class II 
YSOs from different regions. 
The solid black line shows the
best-fitting model while the grey lines represent models which satisfy the above 
mentioned criteria. Among the output parameters, we present the results 
on mass and age estimates in the next section
The assumptions used in the models affect the reliability/accuracy of 
estimated parameters; which are 
considered as inevitable errors which occur in attempts of model 
dependent parameter estimations. 
The estimated parameters are based on the data mentioned above and do not 
include far-IR data.
This introduces relatively large errors in the estimated parameters, 
as suggested by the large number of grey coloured fitted lines
in the SED plots. Since Class I sources peak in mid-IR wavelengths, 
and Class II sources in NIR,  
mass and age estimations based on NIR and MIR are reliable to certain extent.  
In the absence of far-IR wavelengths, disk parameters are less constrained and 
the envelope parameters are highly uncertain. 
Inclusion of far-IR and sub-mm data will reinforce the 
disk nature and characteristics of YSOs. Results presented in this study are 
based on the high resolution
data for a wide sample of candidate YSOs based on combined 2MASS and GLIMPSE data.
The relatively poor resolution due to 
large aperture sizes of 
longer wavelength observations have to be addressed. Inclusion of longer wavelength data 
would effectively reduce the detected 
source density in each region, especially since these are distant star forming regions. 
Hence, instead of a statistical analysis of the properties 
of large number of Class I/Class II sources, it would reduce to individual 
source studies. 

\subsection{Parameters of the identified YSOs}
The number of Class I and Class II sources identified in each region are tabulated
in Table 1, along with the number of IRDCs, Methanol Masers, 
IR bubbles and Star-less clumps.
A rough size of the filamentary star forming region is measured on each 5.8 $\mu$m image 
using ruler option on the SAOIMAGE DS9 image display widget. The approximate 
size (in arc minutes), as well as the distances assumed for each region 
(refer Sec. 3.3) are also tabulated in Table 1.
It is seen that the number of Class I sources roughly scales 
with the number of star-less clumps. 

The estimated mass distributions 
of Class I and Class II sources in the identified 10 regions are shown in Figure 5. 
The histograms 
presented in Figures 5 and 6 may be hugely influenced by an incomplete 
census of YSOs. This study is likely to have missed a number of low mass objects, 
and they are more 
sensitive to (low mass) Class II objects than Class I objects.
It can be seen that Class I sources are found to have a range in 
mass with most of them having mass $\ge$ 8M$_\odot$. 
The upper mass limit in most of the region is
found to be 30 - 32M$_\odot$, though there are a few massive sources.
The most massive YSOs in our study are found in regions 5 and 6. 
Among the Class I sources, 8 sources have estimated mass $>$ 30M$_\odot$; 
4 of them are in region 3, 1 source is in region 4, 2 are in 
region 5 and 1 source in region 6. 
6 of them are in the mass range 30 - 36M$_\odot$; while 2 in 
the 48 - 50M$_\odot$; 
the latter being the most massive ones in this study. These suggest 
that the region studied here are indeed forming massive stars. 
Region 5 has the highest ratio of massive Class I to Class II sources. 
Regions 1 and 2 are found to have more low-mass sources, whereas region 
6 has more sources in the high-mass range.

The mass distribution shows that for all the regions there is 
significant reduction in the 
number of Class II sources with mass higher than 8M$_\odot$; 
while most of Class I sources have masses more than 8M$_\odot$. 
Number of Class I sources in the 
mass range, $<$ 6M$_\odot$ is relatively less; where we can see 
enough number of Class II sources. This may not be due to 
lack of Class I sources with $<$ 6M$_\odot$ mass, 
but due to the detection limit. Observational evidence 
for lower-mass protostars is difficult to obtain. As high-mass protostars 
are much more luminous than lower mass protostars, they are far easier to 
detect. In Class II phase we do find higher fraction of lower mass 
protostars, as it becomes more identifiable in this phase. 
Region 1 is found to have relatively
large number of low-mass Class I sources and it may be due to the fact 
that this region is relatively closer, as being part of the M17SWex. 

The estimated age distributions 
of Class I\&II sources in the identified 10 regions are shown in Figure 6. 
Most of the Class I and Class II sources are with age 
$\le$ 0.1 Myr. In all regions, Class II sources have larger range in age, 
with some sources as old as 3 Myr. 
The age distribution for Class II sources shows that 
in some  regions star formation started at $\sim$3 Myr back, it continued till  
the Class I and Class II are formed together 
in the recent $\le$ 0.5 Myr. Most of the Class I and II soures are found 
to be formed in the last 0.1 - 0.5 Myr. 
Regions 1, 3 and 6 show a more or less continuous formation of 
Class I sources with ages upto $\sim$2.2Myr.

\begin{table*}
\centering
\caption{Filamentary regions and clusters with their source content.}

\begin{tabular}{cccccccccc}
\hline
\hline
&&Distance&&\multicolumn{6}{c}{Number of}\\
Region&Name   &range*&Size &Class I&Class II  &IRDCs&Methanol & IR &Star-less \\
      &   &(kpc)&(')    & & &&Masers& bubbles&clumps\\
\hline
\hline
1 &G14.2-0.55&4-6&20&59& 57 &27&1  &   -&   4 \\
2 &G13.87-0.48&1.9-2.3&14&18& 20 &14&-&  1  &    1 \\
3 &G14.62-0.05 &1.9-2.3&-&71& 54 &40&5  &   3  &    13  \\
4 &G13.26-0.31&4-6&20&17& 5 &16&- &  - &  - \\
5 &G13.05-0.15&4-6&33&163 &32 &40&1 &    1 &$\sim$14 \\
6 &G12.8+0.50&4-6&-&21& 36 &21&1 &    1 &5 \\
7 &G12.34+0.51&4-6&8&9 &11&10& -&  1 &   - \\
8 &G11.86-0.62&4-6&7.5&7 &6 &8&1   &  - & - \\
9 &G11.13-0.13&3.2-3.9&20&34 &11 &20&1 &    1  &   5  \\
10 &G10.67-0.21&4-6&9&26& 9 &20&3  & -  &  1  \\
\hline
\hline 
\end{tabular}
\\
$^*$ as given in the input of SED fitting tool (Refer sec. 3.1) \\
\end{table*}

\par
\subsection{Star formation in Filaments \& clusters}
The description of 8 filamentary star formation sites and 2 candidate clusters of 
Class I and Class II objects including their observed structural details, 
 YSO content, age and mass distributions of YSOs and other sources 
associated are given below. \\

{\bf 1. G14.2-0.55 (Region 1 - Figure 7)}\\
This region is a part of M17SWex which extends $\sim$50 pc southwest 
from Galactic HII region M17. \citet{povich10} 
carried out the census of young stellar content using 2MASS, GLIMPSE, 
MIPSGAL and MSX data and detected $>$ 200 YSOs 
which form B stars. 
For the 64,820 GLIMPSE sources located within a 1$^0 \times 0^0$.75 
field encompassing M17SWex, \citet{kurucz93} reddened stellar 
atmospheres were fitted 
and those sources with poor fit are considered as possible YSOs by 
\citet{povich10}. From these, sources with IR 
excess emission were filtered out using \citet{smith10} color criteria 
and AGB contaminants 
were removed by applying \citet{whitney08} color criteria. 
Assuming M17SWex will form an OB association with a Salpeter IMF, 
they suggested (1) more rapid circumstellar disk evolution in more 
massive YSOs and (2) delayed onset of massive star formation in this region.
We do not study the entire M17Swex region, but a major part of 
filamentary dark clouds, which are termed as region 1 and region 2. 
Our study 
tries to look at the 
general trend in mass and age for YSOs which are being detected in or near 
the filamentary dark clouds.

Distribution of Class I sources 
follow the pattern identical to that of the structure of dark filaments 
giving the 
impression that both are co-located; while most of the Class II 
sources are not closely associated with the filamentary structure. 
Instead of a single elongated 
hub, many tiny filaments are oriented in the decreasing longitude 
direction (away from M17)
giving the appearance of an elongated dark filament. For a distance of 2.1 kpc 
(Povich \& Whitney (2010)) 
we estimate a length of $\sim$12 pc for this filament corresponding to 20'size. 
Class II sources show a peak in mass at around 2 - 3M$_\odot$ and 
Class I sources peak at 8 - 9M$_\odot$. Class I sources also show a 
distribution in 
mass upto 15 - 20M$_\odot$; while Class II sources have masses only 
upto 10M$_\odot$. 
From the age distribution of Class I and Class II sources, 
it can be seen that most of sources are formed at $\le$ 0.1 Myr. 
Some of the Class II sources have ages upto 2 Myr; while most of the Class I sources 
have age $\le$ 0.2 Myr.
The star formation processes are found to have started around 2 Myr ago as evident 
from the presence of almost 10 Class II sources as old as 2 Myr (Figure 6). 
The 
star formation has continued and has produced a maximum number of both 
Class I and Class 
II sources forming together at about 0.1 Myr ago. 
As suggested by \citet{povich10}, 
we also detect an increased formation of YSOs, after a delay of 
$\sim$2 Myr in this region. 

{\bf 2. G13.87-0.48 (Region 2 - Figure 8)}\\
This region is also a part of M17SWex and has been studied by \citet{povich10}. 
For the distance of 2.1 kpc towards 
this region we estimate a length of $\sim$8.5 pc for this filament corresponding 
to the 14' size of the filamentary structure. 
Similar trends in the pattern of Class I and II sources noticed in region 1 
are seen here. The distribution of mass of Class I/II sources 
does not show substantial difference from those noticed in region 1. 
Presence of elongated hubs can be seen here
though the tiny filaments are not obvious. 
Here Class II sources started forming at around 1.6 Myr ago and star formation 
processes reached at its peak in the last 0.1 Myr with most of the Class I and Class II 
forming together. 
Thus we find that region 2 also has older Class II sources, 
but only younger Class I sources. Also Class II sources are found 
to be relatively younger than those in region 1, suggesting that 
the formation in region 2 started later, by about 0.4 Myr. This might support 
the suggestion of sequential 
star formation in the M17 complex. 

{\bf 3. G14.62-0.05 (Region 3 - Figure 9)}\\
Along with IRDCs and starless cores, mid-IR nebulosity is also 
seen in this $l$-$b$ range. We identify this region as a candidate cluster of 
Class I and II 
sources. Highly luminous background emission and absence of definite filamentary 
structure indicate that the
region is more evolved than other regions. Class I sources 
are located mainly in the dark patches. 
Class I sources show a peak in the mass distribution 
at 8 - 13M$_\odot$ and ranges upto 30M$_\odot$. 
The mass distribution of the Class II sources 
peaks at 5 - 6M$_\odot$. No Class I source is seen with mass $<$ 4M$_\odot$. 
Only very few 
Class II sources have mass $>$ 10M$_\odot$; where substantial number of 
Class I sources exist in the 10 - 30M$_\odot$ mass bin.
Both Class I and Class II peak at the age $\le$ 0.1 Myr. 
Some  
of the Class II sources are relatively older, upto 2.1 Myr. 
Similar to the regions 1 and 2,
this region also shows that most of the YSOs are formed about 2 Myr after the 
initial star formation. 
There exist 4 Class I sources with mass $>$ 30M$_\odot$ in this region; with masses 
30.74$\pm$1.67, 33.92$\pm$0.45, 34.49$\pm$1.5 and 35.78$\pm$10.75, (in solar masses) ; 
the last one showing 
the highest error in estimated mass among all YSOs. 

{\bf 4. G13.26-0.31 (Region 4 - Figure 10)}\\
The region consists of a network of tiny filaments, seen inclined to the 
general Galactic plane and parallel to region 5 which is located at a 
higher b value. 
Length of the structure is $\sim$ 20' which stretches from G13.15-0.38 
to G13.4-0.24 and corresponds to a physical length scale in the range 23-34 pc 
for the adopted range of distance (4-6 kpc). 
The increased number of Class I sources above Class IIs 
suggests that this may be 
hosting a very recent star formation site.
All Class I sources have age $\le$ 0.1 Myr; while Class II sources 
are mildly older, up to 0.3 Myr. Hence,
unlike the regions 1 and 2, we detect only recent star formation activity 
in this region.
Most of the Class II sources have mass in the range 3 - 5M$_\odot$. 
More number of Class I sources are seen in 7 - 10M$_\odot$ and 
13 - 14M$_\odot$ mass bins. 
This region also hosts a Class I source with mass 32.4$\pm$4.07M$_\odot$. 
This region is located close to W33, similar to region 5. 
Signatures of HFS with tiny filaments arising parallel to short axis of 
the hub are observed in this region. At some points the filaments are 
entangled together where more number of Class I sources are located. 
The formation of filaments
in this region is likely to be created by the outflows/winds from W33. 
 
{\bf 5. G13.05-0.15 (Region 5 - Figure 11)}\\
This is one of the prominent region which is located near the W33 
complex, where we can see Class I sources closely associated with the 
filamentary structure. 
This is the longest filamentary structure with a size of $\sim$33' 
(corresponding to a physical length in the range 38-56 pc according 
to the distance adopted) located 
to the south-east of W33 complex and is inclined with respect to the Galaxy plane. 
A long strand of dark filament with tiny filaments branching out at some points 
and entangled together at some points gives the impression 
of a hub-filamentary system.

The increased number of 
Class I sources over Class II sources and presence of large number of IRDCs 
along with the morphology 
of the filaments suggest that the region is highly primitive and 
harbors a stellar nursery. Comparatively evolved ultra 
compact HII region W33 
complex should be influencing the structure as well as star formation in this region. 
The Class I sources are seen to be nicely  aligned along the 
length of filaments which extends from G12.8-0.35 to G13.3-0.00  
as seen in Figure 11. 
The mass distribution of Class I sources shows a peak at 9 - 10M$_\odot$. 
Significant fraction of Class I sources have mass $>$ 8M$_\odot$ indicating that 
this region is proliferately forming high mass stars. There are 2 Class I sources 
with mass greater than 30M$_\odot$, with masses 
30.74$\pm$1.06M$_\odot$ and 49.21$\pm$2.82M$_\odot$. Thus this region hosts the 
most massive YSO detected in this study. Only a very few Class II 
sources have mass higher than 10M$_\odot$. It is found that both 
Class I and Class II 
have age $\le$ 0.1 Myr, with some of the Class II sources as old as 0.5 Myr. 
Similar to the
star formation in region 4, this region also has started forming stars 
very recently.   

{\bf 6. G12.8+0.50 (Region 6 - Figure 12)}\\
The region is abundant in Class II sources and dust lane emissions 
compared to other regions and is classified as a candidate cluster.
This is the only region where 
we find significantly more number of Class II sources 
when compared to the number of Class I sources.
Major fraction of Class II sources have mass in the 4 - 6M$_\odot$ range; 
while most of the Class I sources 
have mass upto 16M$_\odot$, with a very few ranging upto 30M$_\odot$. 
Class I sources have age $\le$ 0.1 Myr, where Class II sources are 
as old as 2.5 Myr.
The Class II sources are found to be forming from 2.5 Myr till now, 
whereas the high mass
stars are formed only now. 
The region hosts the second highest massive YSO detected in our study 
with a mass, 48.74$\pm$1.04M$_\odot$. 
The structure and star formation in this region is possibly modified 
by either
the W33 region or the sites of star formation seen in dust emission. 

{\bf 7. G12.34+0.51  (Region 7 - Figure 13)}\\
This region which has relatively faint features compared to other regions, 
consists of a filamentary 
structure with tiny filaments branching out forming a hub-filamentary 
system. The filaments can be seen emerging out of two regions which show 
dust emission, representing the hub.
The filaments are found to be aligned in the same direction. The 8' size of 
this region corresponds to a length scale of 9 - 14 pc for 
the 4-6kpc adopted distance range. 
We detect only a few number of Class I \& II sources in this region. 
Class II sources show a peak in mass in the range 5 - 6M$_\odot$ and 
Class I sources varying upto 16M$_\odot$. Both Class I and Class II sources 
have age $\le$ 0.1 Myr, but some Class II sources are as old as 1.8 Myr. 
As the Class I sources are not found in most part 
of the filaments, 
it appears that the filaments in this region has not attained the critical 
mass to start star formation.

{\bf 8. G11.86-0.62 (Region 8 - Figure 14)}\\
This region has the same pattern as that of region 7. 
Here also signatures of hub-filamentary system can be seen. 
Branching out of filaments to form a network is clearly observable in this region. 
Here also for the adopted distance range, we estimate a length of 
8.7 - 13 pc corresponding to the 7.5' size of the filamentary structure.
There are no Class I sources with mass $<$ 7M$_\odot$ and most 
of them have in the 
range 8 - 9M$_\odot$. Masses of Class II sources vary from 2 - 11M$_\odot$. 
Most of the 
Class I and Class II sources have age $\le$ 0.1 Myr. 
A few of the Class II have age 2.7 - 2.9 Myr. 

{\bf 9. G11.13-0.13 (Region 9 - Figure 15)}\\ 
This is one of the well-studied filamentary IRDC.
\citet{carey98} confirmed the presence of dense molecular gas in this 
cloud using millimeter spectral lines of H$_2$CO. 
\citet{carey00} presents the 850 $\mu$m and 450 $\mu$m continuum images from 
SCUBA observations and postulates the 
presence of early stages of star formation within the cloud. 
\citet{john03} demonstrated the underlying radial structure of 
the cloud using 850 $\mu$m
observations and according to their findings this is the first molecular 
filament observed to have a 
radial profile similar to that of a non-magnetic isothermal cylinder. 
\citet{henning10} made use of Herschel data to detect the embedded 
population of pre- and protostellar cores in this. 
Out of 18 cores they characterized using SED analysis, two are of with 
masses over 50M$_\odot$, implying the presence of massive star formation.  
The presence of high mass star formation 
in the cloud has been noted by other authors also; \citet{chen10} 
(presence of EGOs as a tracers), 
\citet{pillai06} (water and methanol emission), \citet{gomez11} 
(using methanol emission). 
 
A distinct filamentary cloud structure with tiny filaments originating 
at some points is seen here. It is interesting to see that Class I 
sources are aligned along the string of filamentary structure; while 
Class II sources are located more or less away from the main filament.
The approximate size of the filament is $\sim$20'. This corresponds to a 
length of $\sim$21 pc for the distance 3.6 kpc towards this region.
Most of the Class I sources have mass 
$<$ 10M$_\odot$ and age $\le$ 0.1 Myr with a peak 
distribution in mass at 9 - 10M$_\odot$ and a very few ranges upto 30M$_\odot$. 
Many Class II sources have mass in the range 5 - 9M$_\odot$ and age 
ranges to 0.5 Myr. 
Many IRDCs, star-less clumps (5 with near distance solution), 
one maser source and one IR bubble are seen 
in this location. IRDCs are also 
nicely aligned in these filaments. 
Most of the star formation in this region is found to be in the last 0.5 Myr.

{\bf 10. G10.67-0.21. (Region 10) (Figure 16)}\\
Tiny filaments emanating from smaller round hubs are grouped to make a 
clustered form of filaments in this region. This region with a 9' size 
corresponds to a length scale in the range 10 - 16 pc for the 
adopted distance range of 4 - 6 kpc. 
Most of the Class I and Class II have the age $\le$ 0.1 Myr. 
Masses of Class I sources 
vary upto 26M$_\odot$ with many peaking at 8 - 9M$_\odot$. Class II sources 
have mass 
in the range 2 - 13M$_\odot$. 
This is a site experiencing star formation in the last 0.5 Myr.

\section{Discussion} 
We have studied YSOs in 10 star forming regions in the inner Galaxy. 
The presence of massive Class I
sources along with other tracers of high mass star formation prove 
that these are sites
of high mass star formation. The regions show filamentary dust lanes 
and Class I sources are found to be 
preferentially located along them. As these regions are located near 
the well-known HII regions, the structure and
star formation in these regions may be dictated by them. 

Figure 17 shows the location of tracers of 
high mass star formation such as 
IRDCs, masers, star-less clumps, 
IR bubbles/HII regions, radio sources, EGOs and HMPOs 
identified in the galactic 
region of our study. Though the 10 regions studied here are selected 
based on the density of the
Class I and Class II sources, a few other concentrations of above mentioned 
tracers can be identified in this plot. Thus, figure 17
shows the location of these concentrations and suggests that there may be a 
few more high mass star forming sites, close to
the regions studied here. We do not study them as we detect only a few Class 
I sources in these regions. It is possible that 
these regions are in the very early stage of star formation. 
Presence of such regions 
once again strengthens the arguement that this region of the Galaxy 
is undergoing high mass star formation, which are potential targets 
for further studies.

Out of the 
10 regions studied here, 4 regions (regions 1, 2, 3 and 6) show 
signatures of delayed 
star formation. Though our
results suggest a delay in the formation of Class I sources (massive stars), 
this is
only indicative because of the incompleteness in the data of Class II sources.
Regions 1, 2 and 3 are located close to M17SWex. 
There is indication of 
sequential star formation in this region due to the effect of M17SWex. 6 regions 
(regions 4, 5, 7, 8, 9 and 10) were found to be forming the high and 
low mass stars together. Stars are being formed in these regions in 
the last 0.5 Myr. 

Figure 18 shows the plot of age vs mass for the Class I and Class II 
sources with estimated errors in masses and ages. 
In the plots we have truncated the limits in 
both X and Y axes, 
including most of the sources; while excluding the extreme ones. 
The interpretations of this plot are likely to be affected by the incomplete 
number of observed YSOs in the region. Also,  
errors on many of the masses and ages are greater than the predicted 
values themselves. 
In the case of Class I sources, the plot is shown only upto 1 Myr, 
though there are 13 sources with ages more than 1 Myr and upto 8 Myr 
having mass less than 8M$_\odot$. It is possible that these older 
Class I sources may be Class II 
sources viewed edge-on, through their disks.
Most of the Class I sources detected in the regions 
are found to be younger than 0.5 Myr and this plot helps to understand 
the mass distribution of this population. 

The mass of Class I sources were found to be between 2 - 32M$_\odot$, with 
most of the sources more massive than 8M$_\odot$. Thus, high
mass stars are being formed in these regions in the last 0.5 Myr. It can be
seen that the relatively low mass Class I sources have age upto 0.5 Myr, 
whereas the more massive sources are younger than 0.2 Myr. It is well 
known that massive stars form much quicker than 
their low-mass counterparts. One would therefore not expect to see 
massive Class I sources with ages greater than a few tenths of a 
Myr. The left-hand plot in Figure 18 supports this theory.
The Class II sources are 
formed in the age range 0 - 3 Myr, but most of the sources
have ages $\le$0.6 Myr. There are about 20 sources which are older than 3 Myr, 
up to 10 Myr and having masses less than 5M$_\odot$. This suggests that most 
of the Class II sources have similar age range as the Class I sources. 
The mass range of the Class II sources is found to be 0.5 - 20 M$_\odot$, but 
most of the sources have mass less than 10 - 12M$_\odot$. Most of the older 
Class II sources were found to have mass less than 8M$_\odot$, except a 
few massive sources. The above numbers are only suggestive and could 
be affected by the incompleteness in the data. We do still find that many - in 
some regions most - of the Class II sources appear to be as young as the 
Class I sources. This can be either due to observational 
bias towards massive YSOs which evolve more rapidly than their 
low-mass counterparts or some of the class II sources are mis-identified as 
Class I in the colour-colour analysis because of their 
orientation (i.e., being viewed pole-on). The environment around each sources, 
orientation 
of the source with respect to the observer and the mass of the central 
object can alter the observed differences in colours and SEDs of Class I 
and Class II evolutionary stage objects thereby affecting the 
properties estimated for them and can be the reason for 
obtaining large population of very young Class II sources. Apart from these
biases in the classification of the YSOs, this study supports the 
evolution of Class I sources into Class II sources. 

In summary, we detect YSOs in
the mass range 0.5 - 31M$_\odot$ and $\sim$0.1 - 3 Myr age range, with most of 
them in the $\le$0.5 Myr age range. These regions, thus prove to be abundant 
in massive YSOs in the early stages of formation, pointing to their 
potential for further studies.
One of the regions studied here is found to have a large number of
high mass Class I sources, suggesting that this region (region 5) is 
forming a massive and rich cluster of high mass stars. 

We observe 
HFS in most of the early star forming complexes of filamentary IRDCs. 
In most cases elongated hubs are seen with tiny filaments. 
We see different orientations 
of filaments with respect to the central hub, like, the filaments 
parallel to hub, filaments radiating in different directions, and in some cases 
filaments are entangled. \citet{myers09} explains the different models of 
formation mechanism of HFS. Analysis of each filamentary star formation sites 
and their environment can lead to the understanding of formation scenario of HFS.
The filamentary regions are located in the vicinity of massive HII regions, 
suggesting that the
filamentary features could be created due to these nearby HII regions. 
Our regions are located at varying
distances from the HII regions and some are located relatively far away. 
Also, some of these regions are
found to host star formation upto 3 Myr ago, with the formation of low mass stars. 
Thus, the early formation of small number of low mass stars is not 
found to destroy the filaments. 

Studies of star formation in filamentary structures is an upcoming area of 
scientific interest as the new space missions like Herschel and other 
sub millimeter observations have recently shown the complex systems of 
filamentary structures. These informations have revolutionized the previous
theory 
of star formation that only self accreting molecular clouds of circular size 
take part in star formation. 
We have shown that the GLIMPSE data \& images are enough to 
identify filamentary structures and thereby contribute in understanding the 
origin and geometry of early star forming processes.

\section{Conclusion} 
Major findings of our study can be summarized as follows: 
\begin{itemize}
\item We have identified 1107 Class I and 1566 Class II sources in the galactic
region $10^0 < l < 15^0$ , $-1^0<b<1^0$. 
\item  We have identified 8 early star forming sites of 
filamentary structures and 2 candidate 
clusters of Class I and Class II YSOs. Class I sources are closely 
associated with the 
infra red dark filaments; while Class II sources are located randomly 
in these regions. All these are found to be co-located with other 
high mass star forming tracers. 
\item In all the regions identified, the observed Class I sources are with 
age $\le$ 0.5 Myr, while Class II sources 
have ages in the range $\sim$0.1 - 3 Myr. Majority of the Class I sources 
are  $\ge$ 8M$_\odot$; while Class II sources in the 
0.5 - 10 M$_\odot$ mass range. Low mass objects are incomplete 
in this group due to low flux levels.
\item 4 regions studied are found to show a delay in the formation of most of the YSOs. 
Our analysis supports the sequential 
star formation 
of M17SWex complex as suggested previously.
\item Filamentary and hub-filamentary features are found in most of these regions, 
which harbor star formation in the $\sim$0.1 - 3 Myr age range. The length of 
the identified filaments are estimated as 8'-33' ($\sim$ 9 - 56 pc).  
\item This study brings to focus 10 sites of massive star 
formation in the inner Galaxy, harboring very young YSOs. 
We suggest that these are
potential targets to understand the formation and evolution of massive YSOs.
\end{itemize}
 
\section{acknowledgments}
{We thank the anonymous referee for the useful comments and
suggestions which helped in improving the paper. 
BB gratefully acknowledges University
Grants Commission, New Delhi, for financial support through RFSMS Scheme
and Indian Institute of Astrophysics, Bangalore, 
for hospitality, where most of this work was done.
This research has made use of the data products from the
GLIMPSE survey, which is a legacy science program of the Spitzer 
Space Telescope funded
by the National Aeronautics and Space Administration 
and also the data products from Two Micron All Sky Survey (2MASS),
which is a joint project of the University of Massachusetts and the Infrared Processing
and Analysis Center/California Institute of Technology, funded by the National
Aeronautics and Space Administration and the National Science Foundation.}

%%%%%%%%%%%%%%%%%%%%%%%%%%%%%%%%%%%%%%%%%%%%%%%%%%%%%%%%%%%%%%%%%%%%%%%%  
\clearpage
%%%%%%%%%%%%%%%%%%%%%%%%%%%%%%%%%%%%%%%%%%%%%%%%%%%%%%%%%%%%%%%%%%%%%%%%%%

%% FIGURE - 1%%
\begin{figure*}
\begin{center}
\subfigure{
\includegraphics[width=7cm]{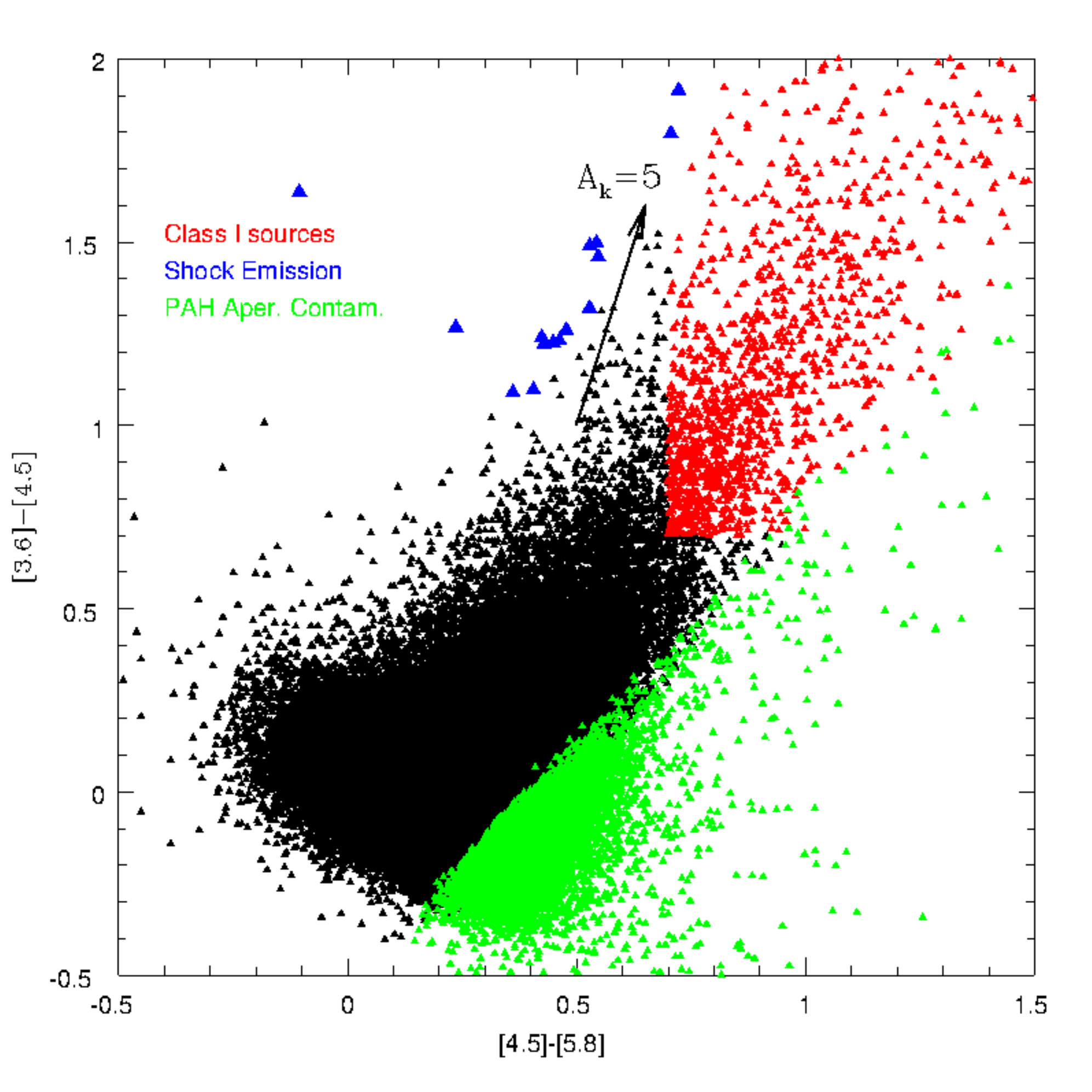}
        }
\subfigure{
\includegraphics[width=7cm]{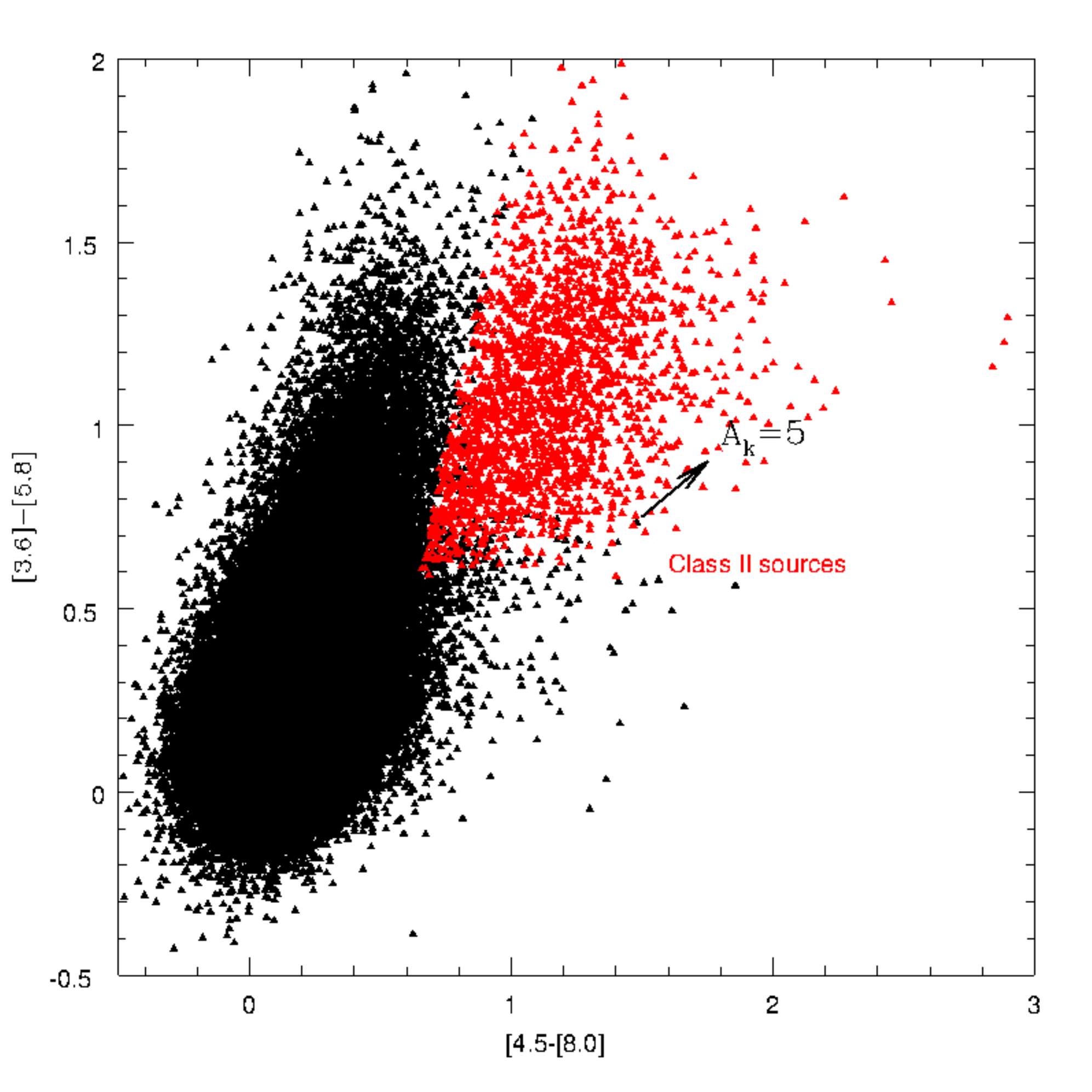}
        }
\end{center}
\caption{IRAC color-color Diagrams showing YSOs and non-YSO sources. {\sl Left:} 
Shock emission sources as blue triangles, PAH aperture contaminants as green triangles and  
Class I sources as red triangles. {\sl Right:} Class II sources as red triangles. Unclassified 
sources as shown as black triangles in both figures. Extinction vector corresponding to A$_k$=5 
mag is shown.}
\end{figure*}

%% FIGURE -2 %%
% Class I & Class II distribution

\begin{figure*}
\centering
\includegraphics[trim=3cm 0.5cm 2cm 0cm, clip=true, totalheight=0.35\textheight, angle=0]{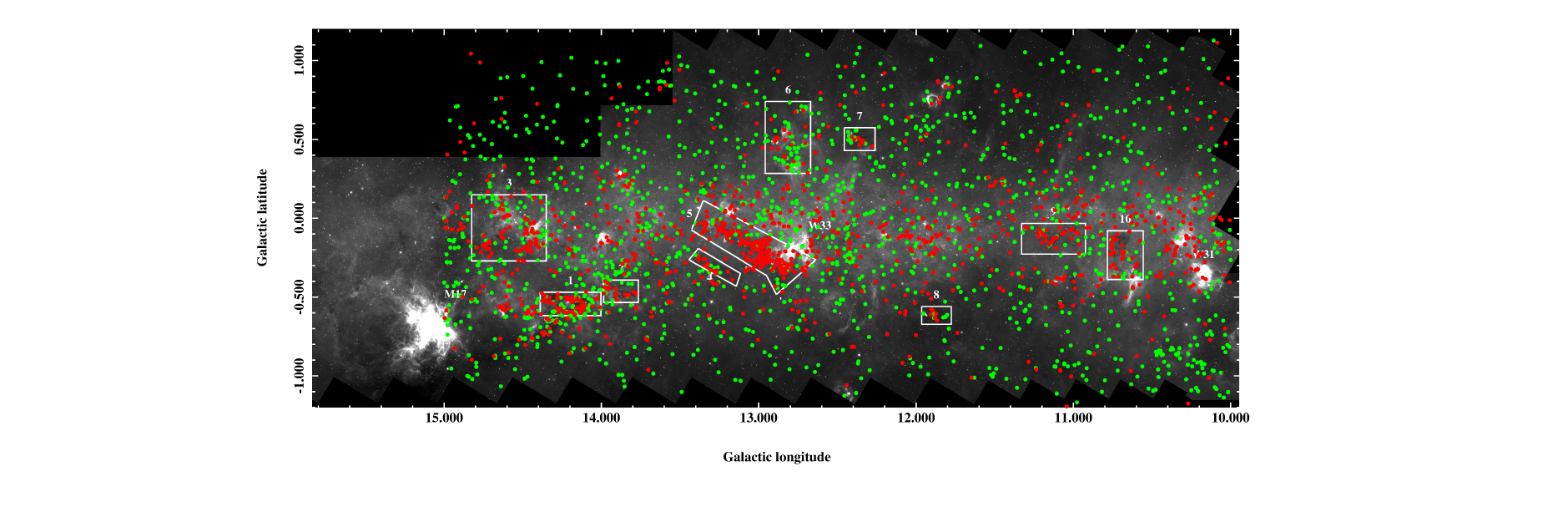}
\caption{The region of our study as seen in the GLIMPSE 5.8$\mu$m image. 
X-axis is Galactic longitude, $l$ and Y-axis is Galactic latitude, $b$. 
Distribution of Class I sources (red points) \& Class II sources (green) in the region are shown. 
Regions marked inside the white boxes are the 
identified star formation sites. Location of M17, W33 and W31 are also shown.}
\end{figure*}

%% FIGURE - 3 %
% Histogram: IRDCs-YSOs
\begin{figure*}
%\epsfxsize=7cm
%\epsffile{fig3.pdf} 
\includegraphics[width=7cm]{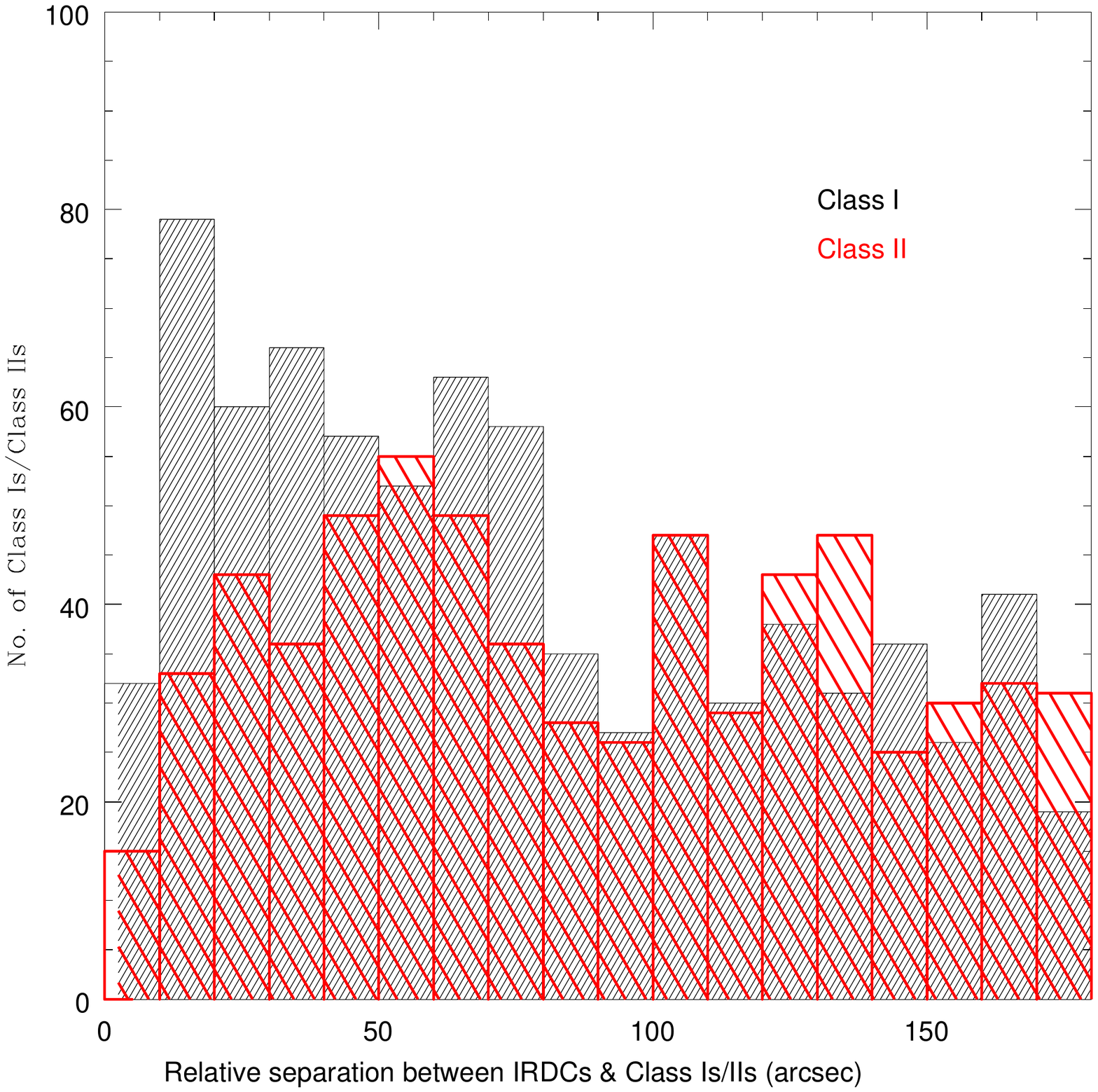}

\caption{Histogram showing the relative separation of IRDC centres and Class I/II 
sources. More number of Class I sources are associated with IRDCs than Class IIs 
for smaller angular separations.}
\end{figure*}

%% FIGURE -4 %%
% SEDs
\begin{figure*}
\begin{center}
\subfigure{
\includegraphics[width=5cm]{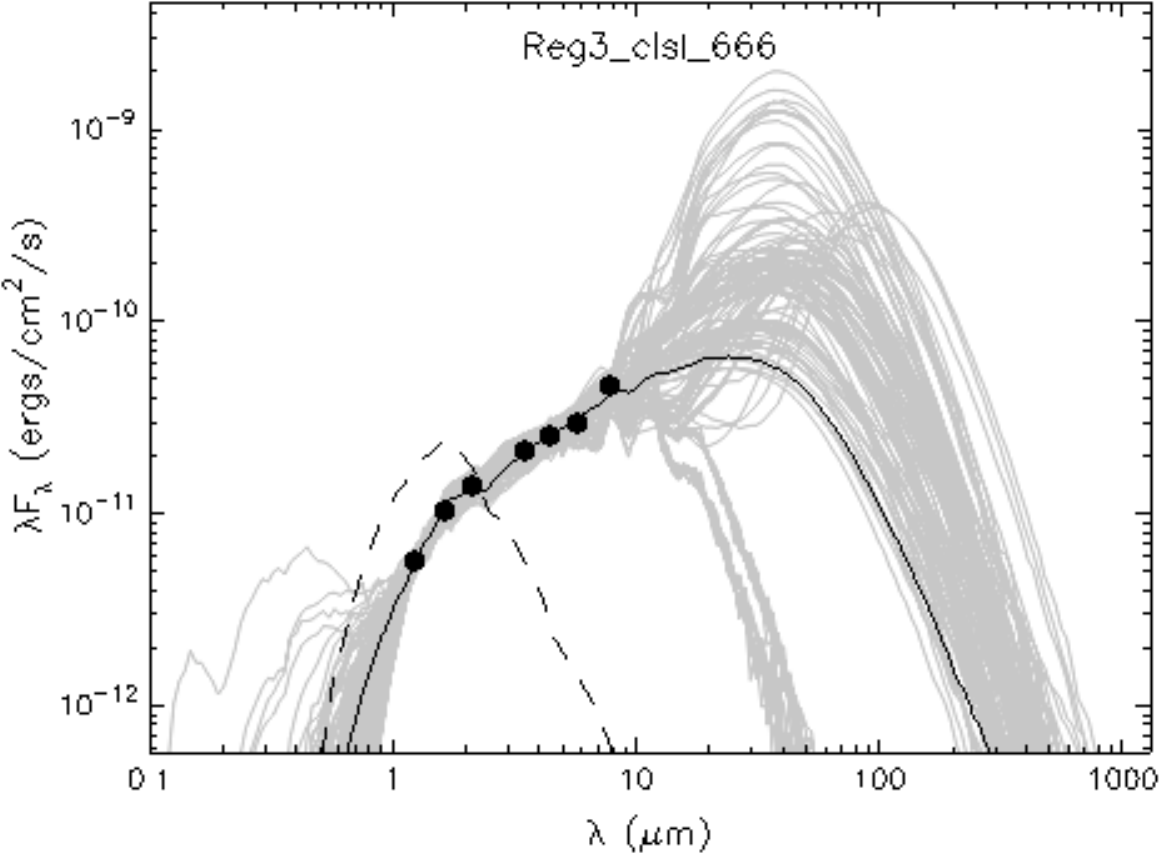}
        }
\subfigure{%
\includegraphics[width=5cm]{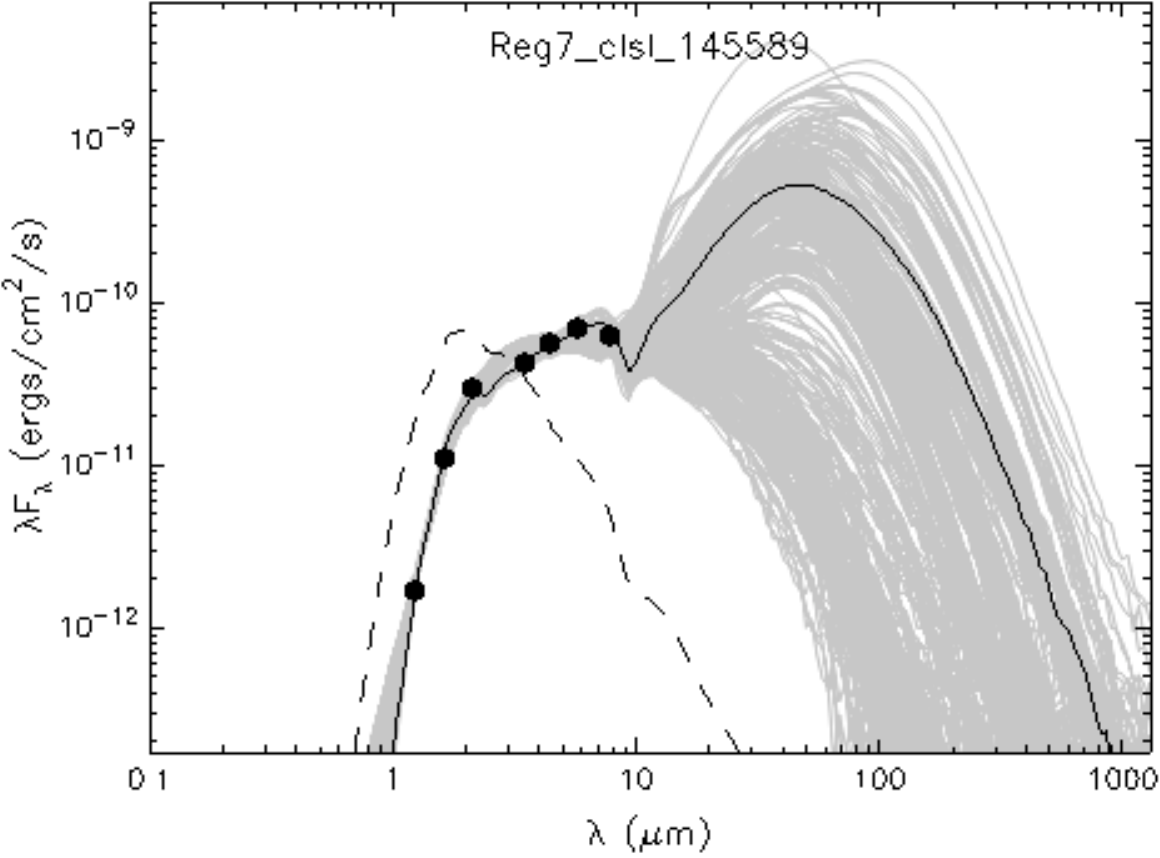}
        }\\
\subfigure{
\includegraphics[width=5cm]{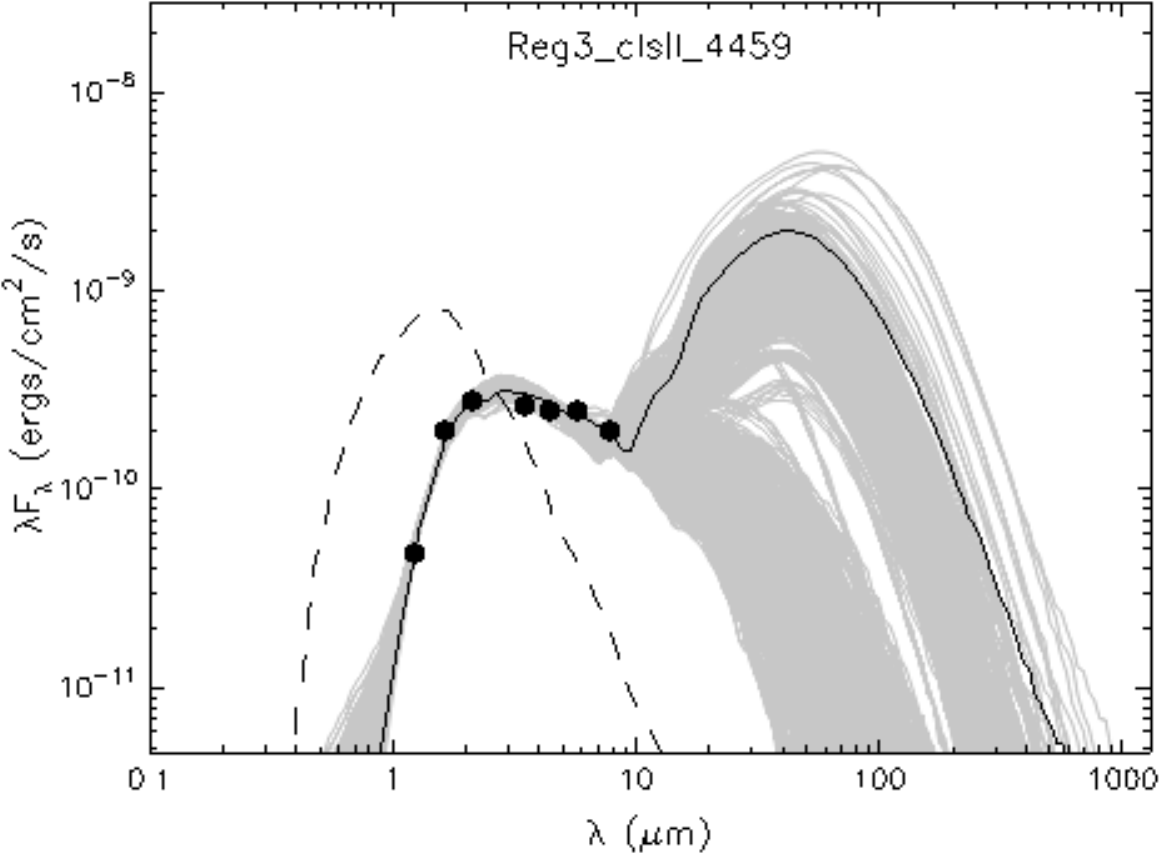}
        }
\subfigure{%
\includegraphics[width=5cm]{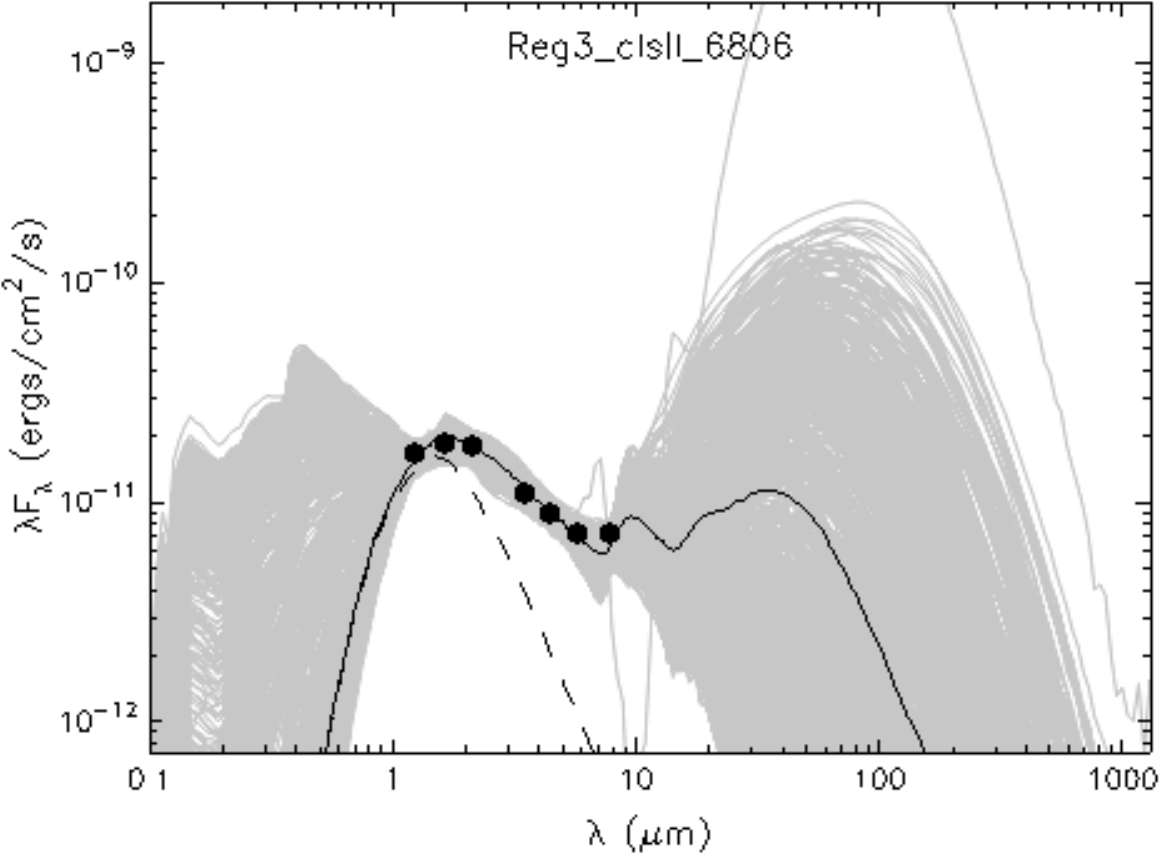}
        }\\
\end{center}
\caption{SEDs of Class I and Class II sources constructed using 
$JHK_s$ and IRAC magnitudes. The top 2 sources are the Class I 
sources from region 3 and region 7 with ages 0.004 $\pm$ 0.068 Myr, 
0.014 $\pm$ 0.075 Myr and 
masses 2.09 $\pm$ 0.3M$_\odot$, 7.14 $\pm$ 0.04M$_\odot$ respectively. 
Bottom sources are Class II sources from region 3 with ages 
0.469 $\pm$ 0.074 Myr, 
0.013 $\pm$ 0.041 and masses 4.89 $\pm$ 0.03M$_\odot$, 
10.4 $\pm$ 0.05M$_\odot$ respectively.}
\end{figure*}

%% FIGURE -5 %%
\begin{figure*}
%\epsfxsize=17cm
%\epsffile{fig5.pdf} 
\includegraphics[width=17cm]{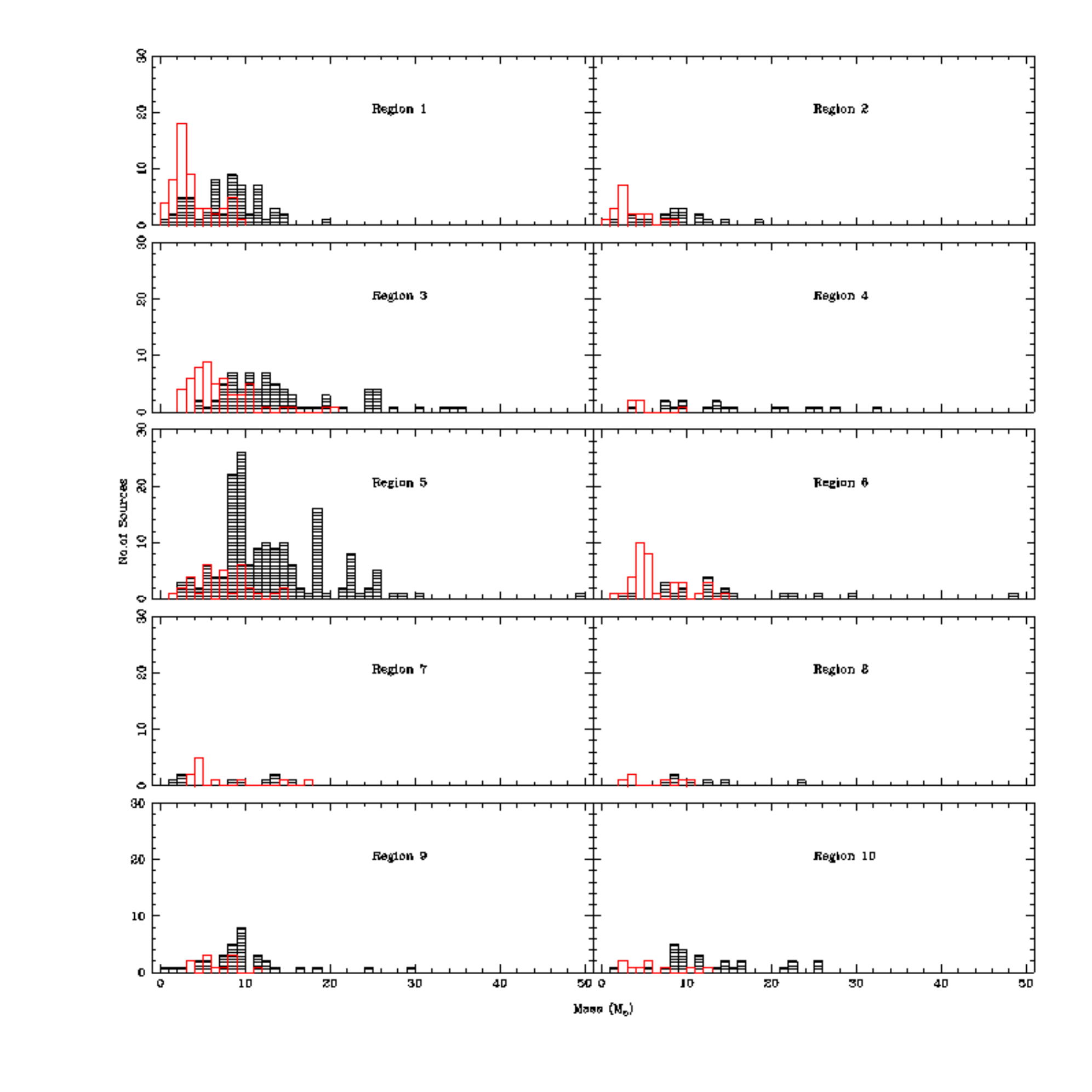}
\caption{Mass distribution of Class I sources (black shaded blocks) and 
Class II sources (red blocks) identified in 10 regions. }
\end{figure*}

%% FIGURE-6%%
\begin{figure*}
%\epsfxsize=17cm
%\epsffile{fig6.pdf} 
\includegraphics[width=17cm]{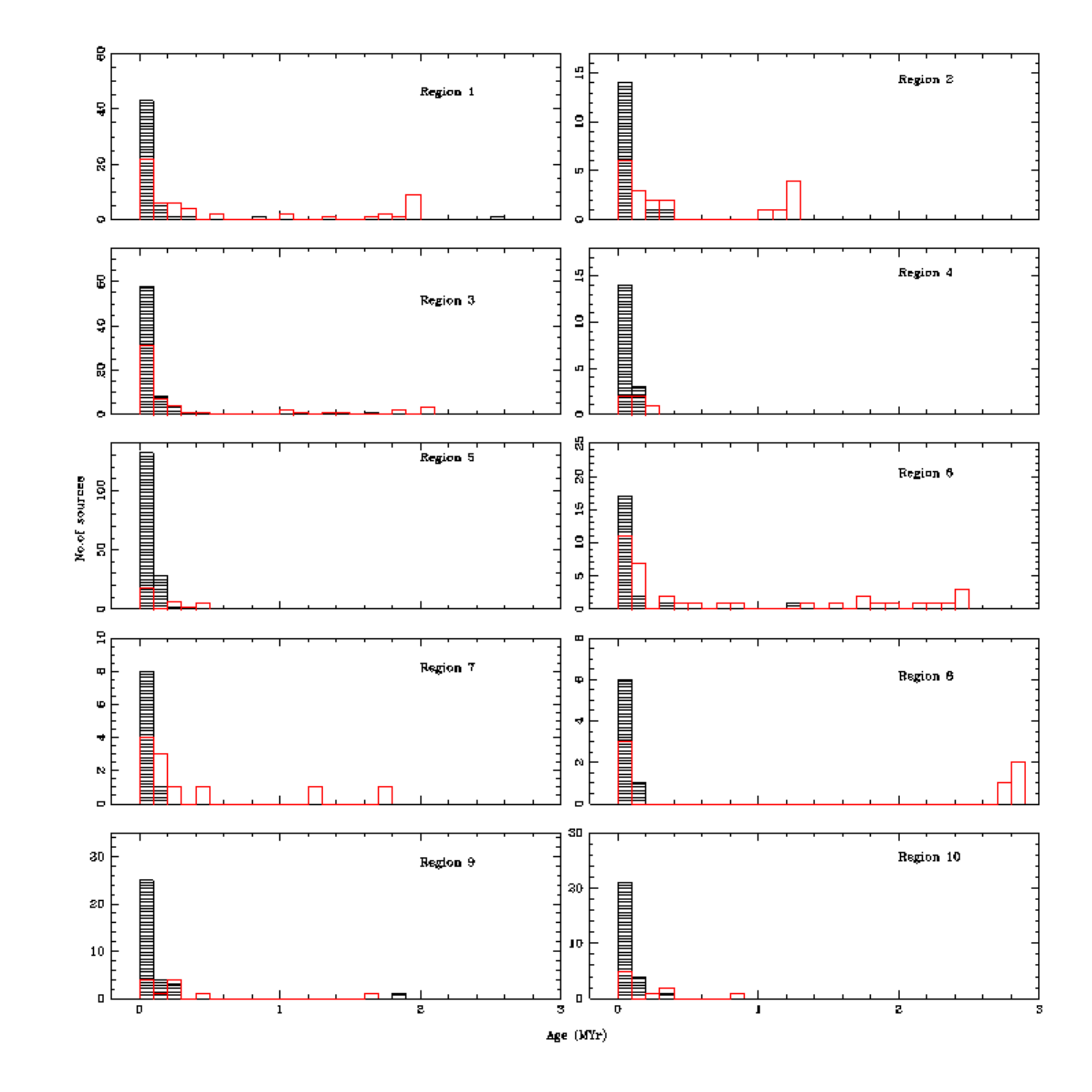}
\caption{Age distribution of Class I sources (black shaded blocks) and 
Class II sources (red blocks) identified in 10 regions. }
\end{figure*}

%% FIGURE 7 ..%%
\begin{figure*}
\centering
\includegraphics[trim=1cm 3cm 2cm 3cm, clip=true, angle=0]{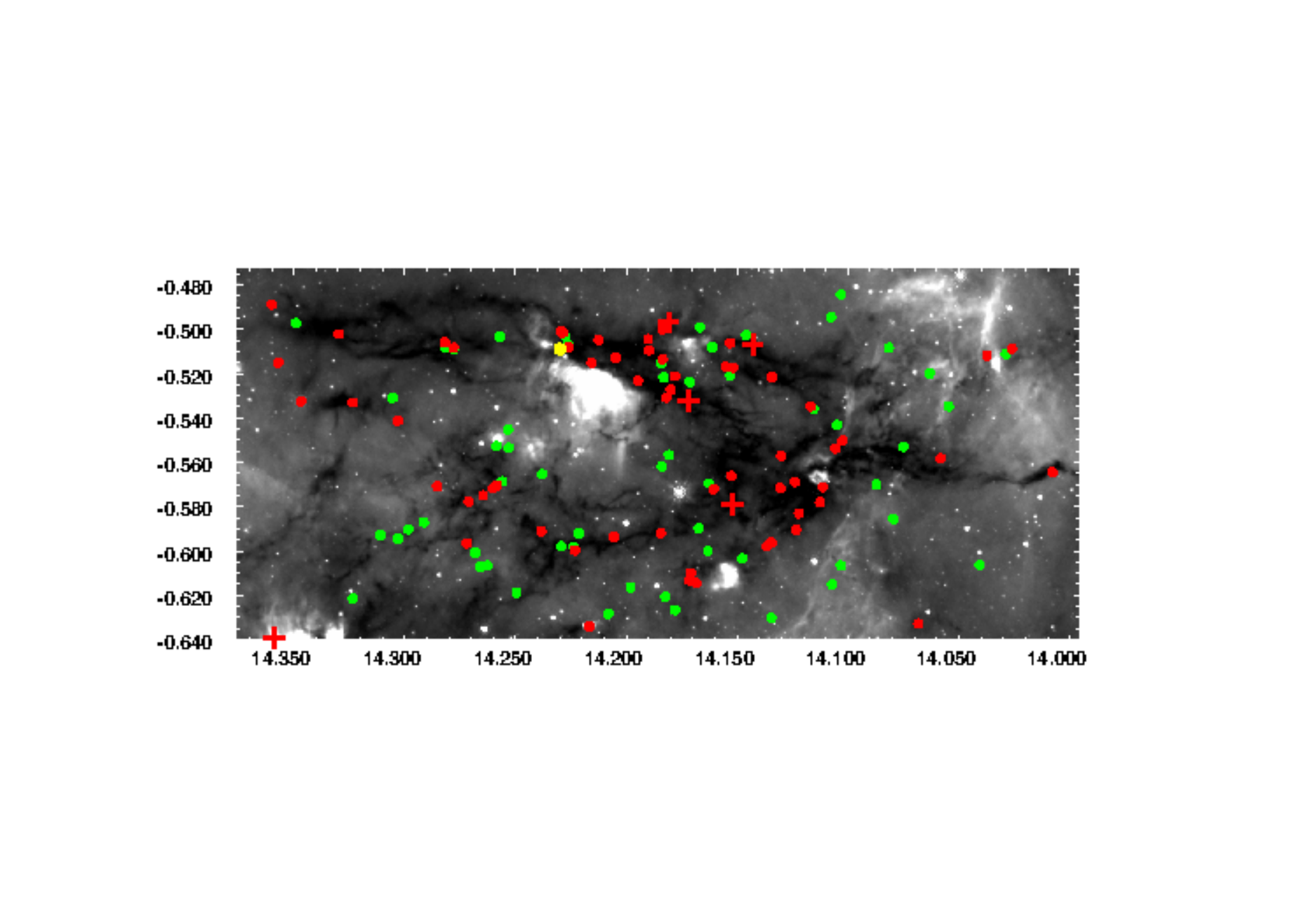}
\caption{G14.2-0.55 (Region 1). Class I sources are shown in red circles and Class II sources in green 
circles. Only those YSOs considered for SED fitting are shown. Location of 
tracers like, Masers, star-less clumps are also marked. Yellow:Masers (without 
distance information), 
Blue:Near 3-Kpc arm Masers, Magenta:Far 3-Kpc arm masers, 
Red crosses: Star-less clumps with near distance solutions assumed, Red diamonds: 
Star-less clumps with far solutions.}
\end{figure*}
%% FIGURE 8 ..%%
\begin{figure*}
\centering
\includegraphics[trim=1cm 2cm 2cm 2cm, totalheight=0.3\textheight, clip=true, angle=0]{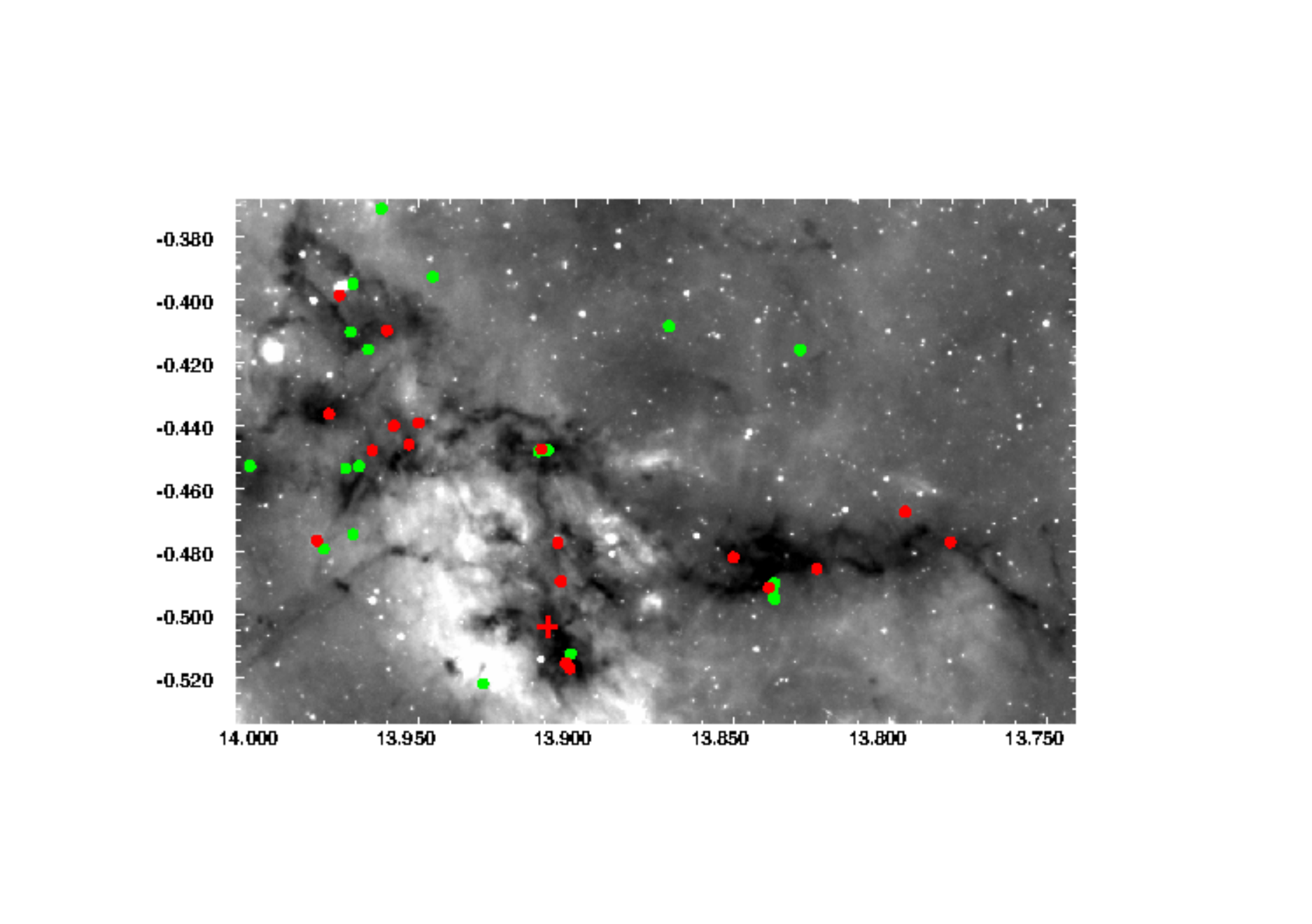}
\caption{G13.87-0.48 (Region 2)}
\end{figure*}
%% FIGURE 9 ..%%
\begin{figure*}
\centering
\includegraphics[trim=0cm 0cm 0cm 0cm, totalheight=0.4\textheight, clip=true, angle=0]{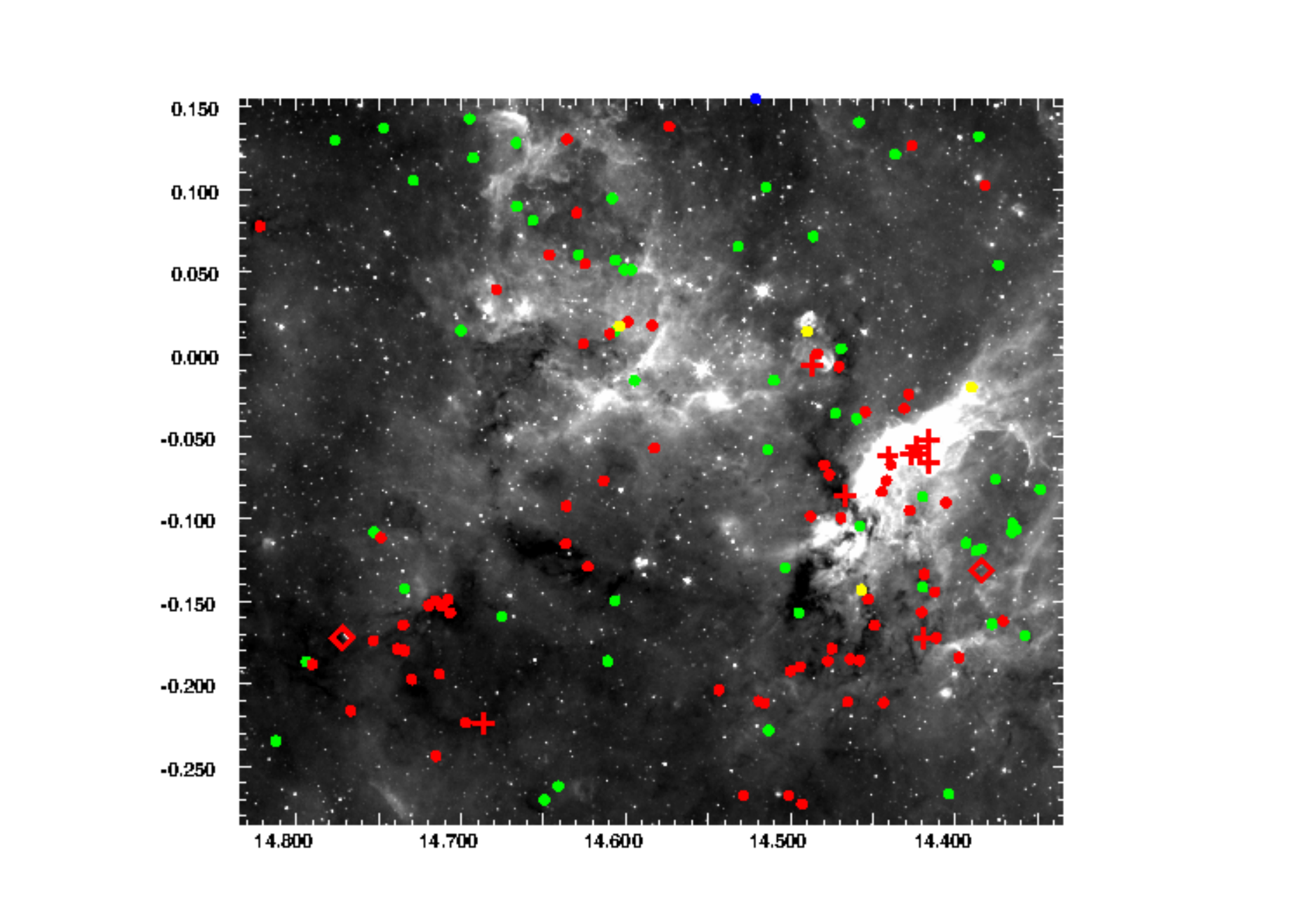}
\caption{G14.62-0.05 (Region 3)}
\end{figure*}
%% FIGURE 10 ..%%
\begin{figure*}
\centering
\includegraphics[trim=1cm 1cm 1cm 0.5cm, totalheight=0.4\textheight, clip=true, angle=0]{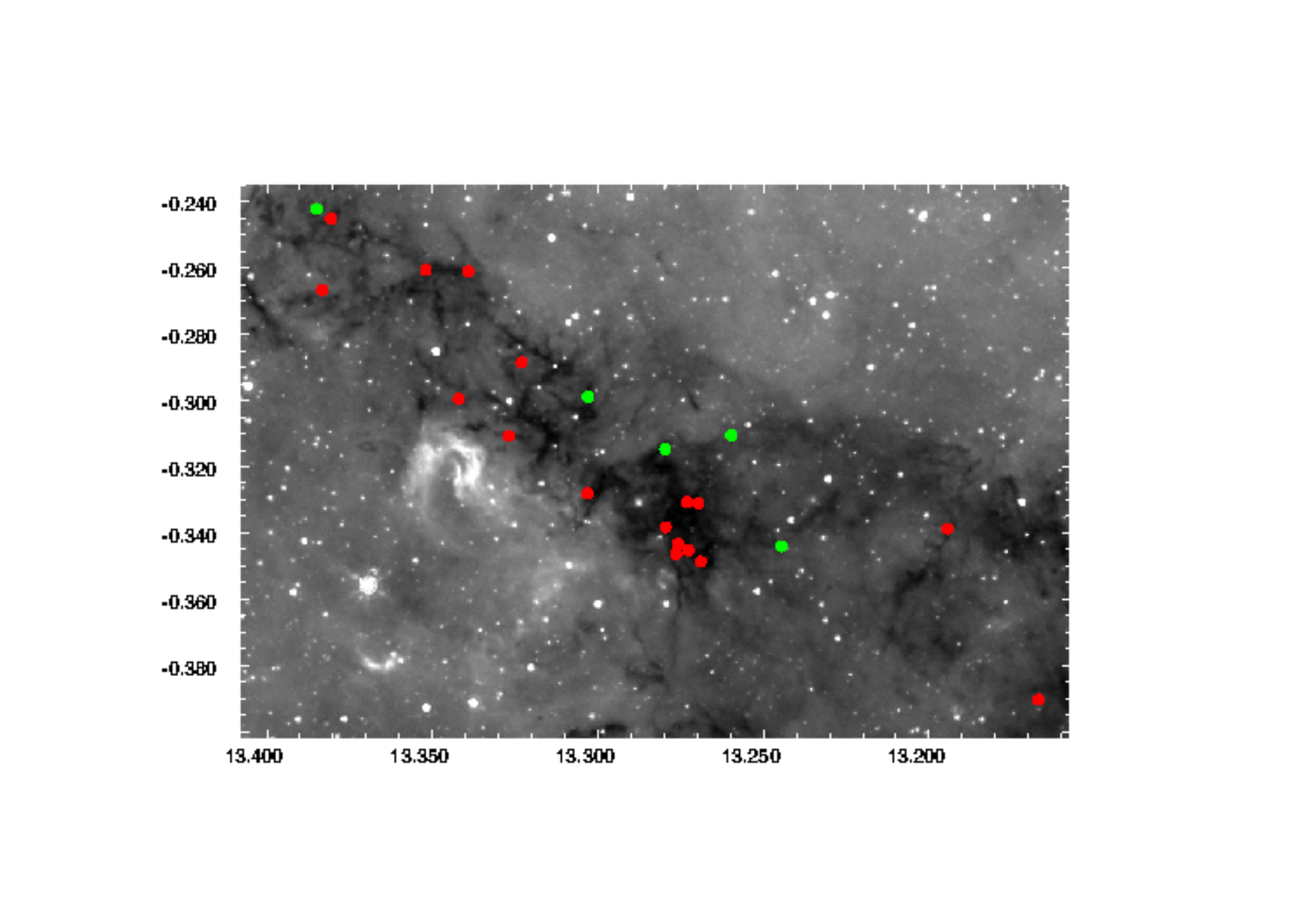}
\caption{G13.26-0.31 (Region 4)}
\end{figure*}
%% FIGURE 11 ..%%
\begin{figure*}
\centering
\includegraphics[trim=0cm 0cm 0cm 0cm, totalheight=0.4\textheight, clip=true, angle=0]{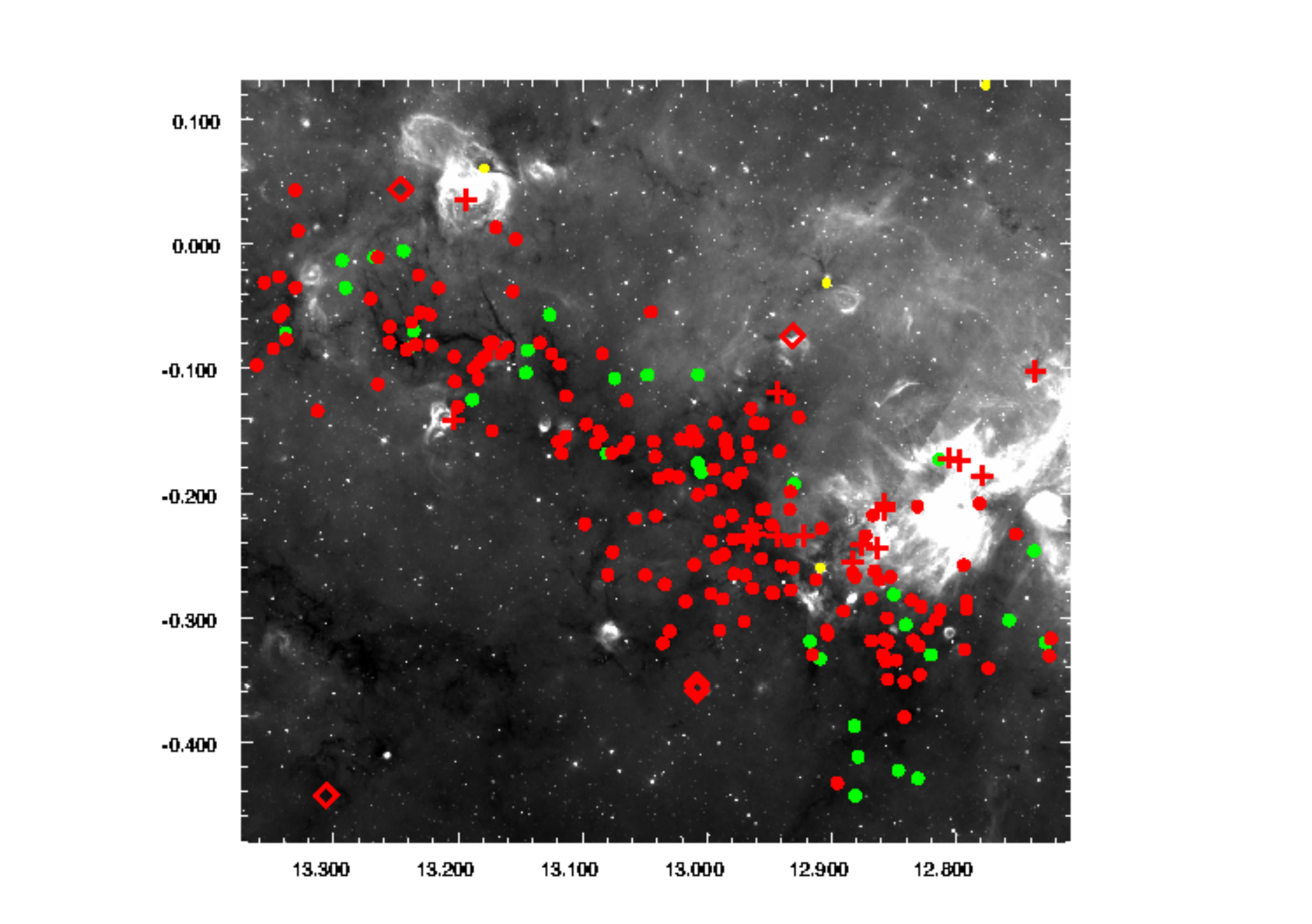}
\caption{G13.05-0.15 (Region 5)}
\end{figure*}
%% FIGURE 12 ..%%
\begin{figure*}
\centering
\includegraphics[totalheight=0.4\textheight]{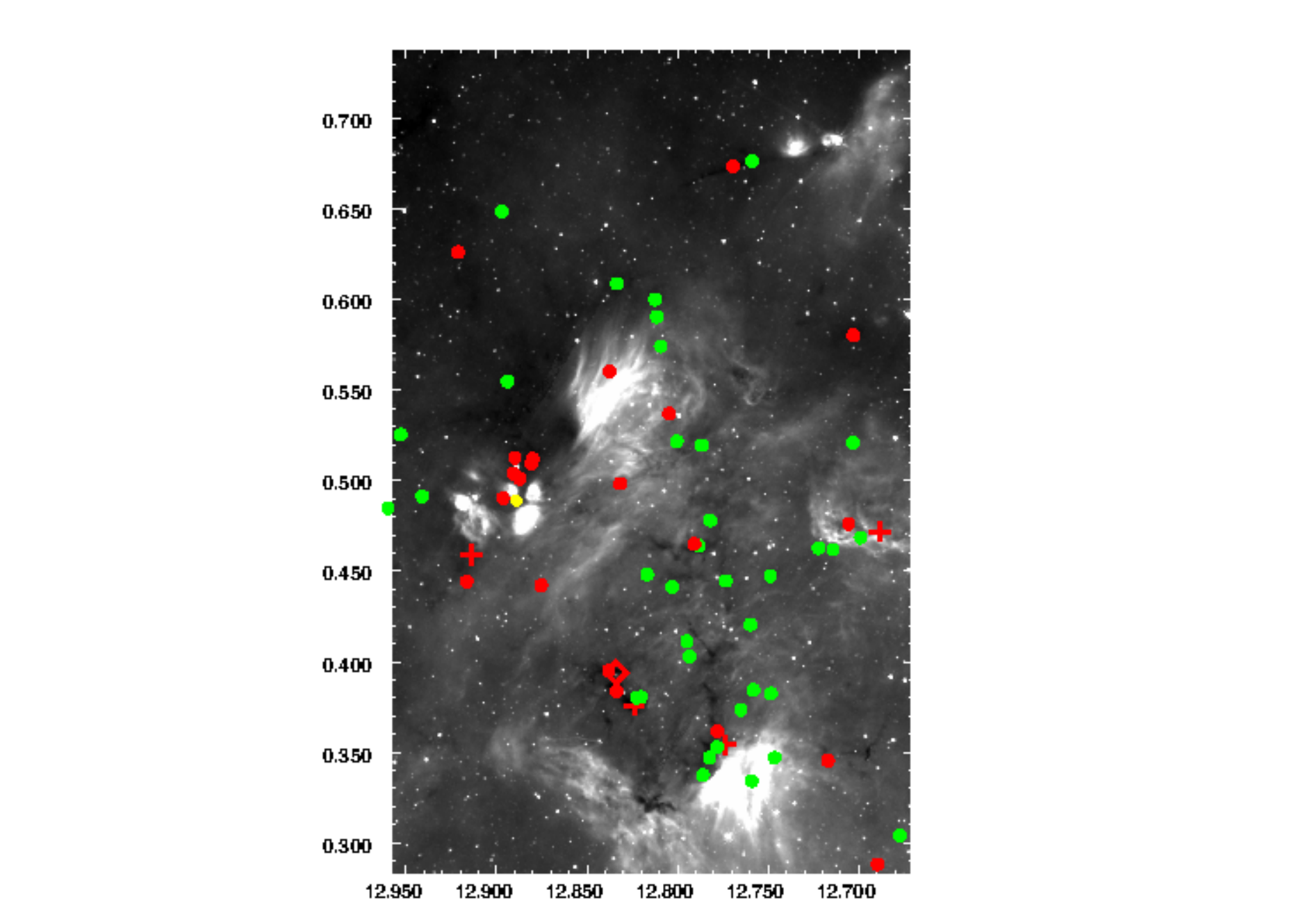}
\caption{G12.8+0.50 (Region 6)}
\end{figure*}
%% FIGURE 13 ..%%
\begin{figure*}
\centering
\includegraphics[trim=0cm 1cm 0cm 2cm, totalheight=0.3\textheight, clip=true, angle=0]{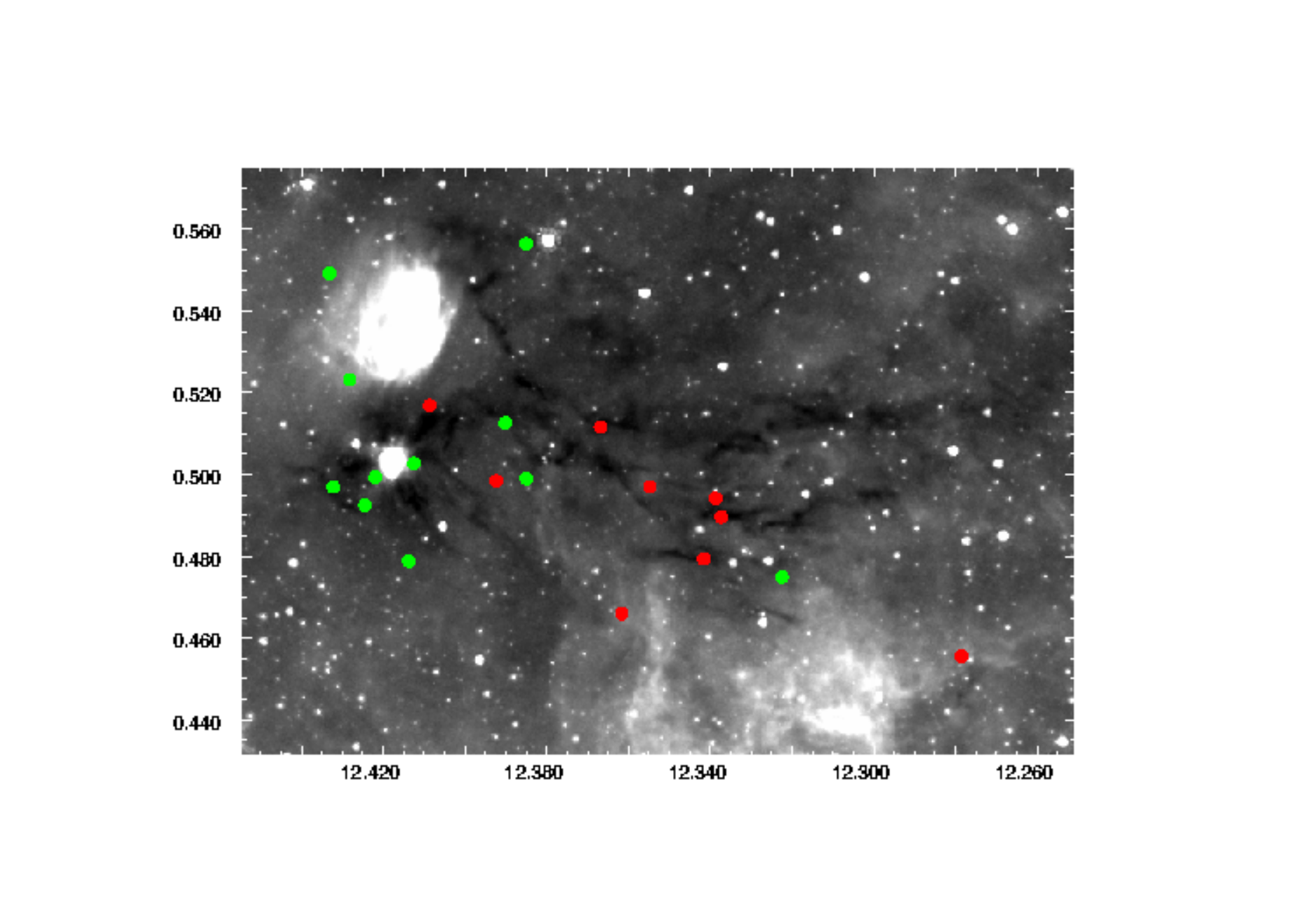}
\caption{G12.34+0.51 (Region 7)}
\end{figure*}
%% FIGURE 14 ..%%
\begin{figure*}
\centering
\includegraphics[trim=0cm 1cm 0cm 2cm, totalheight=0.3\textheight, clip=true, angle=0]{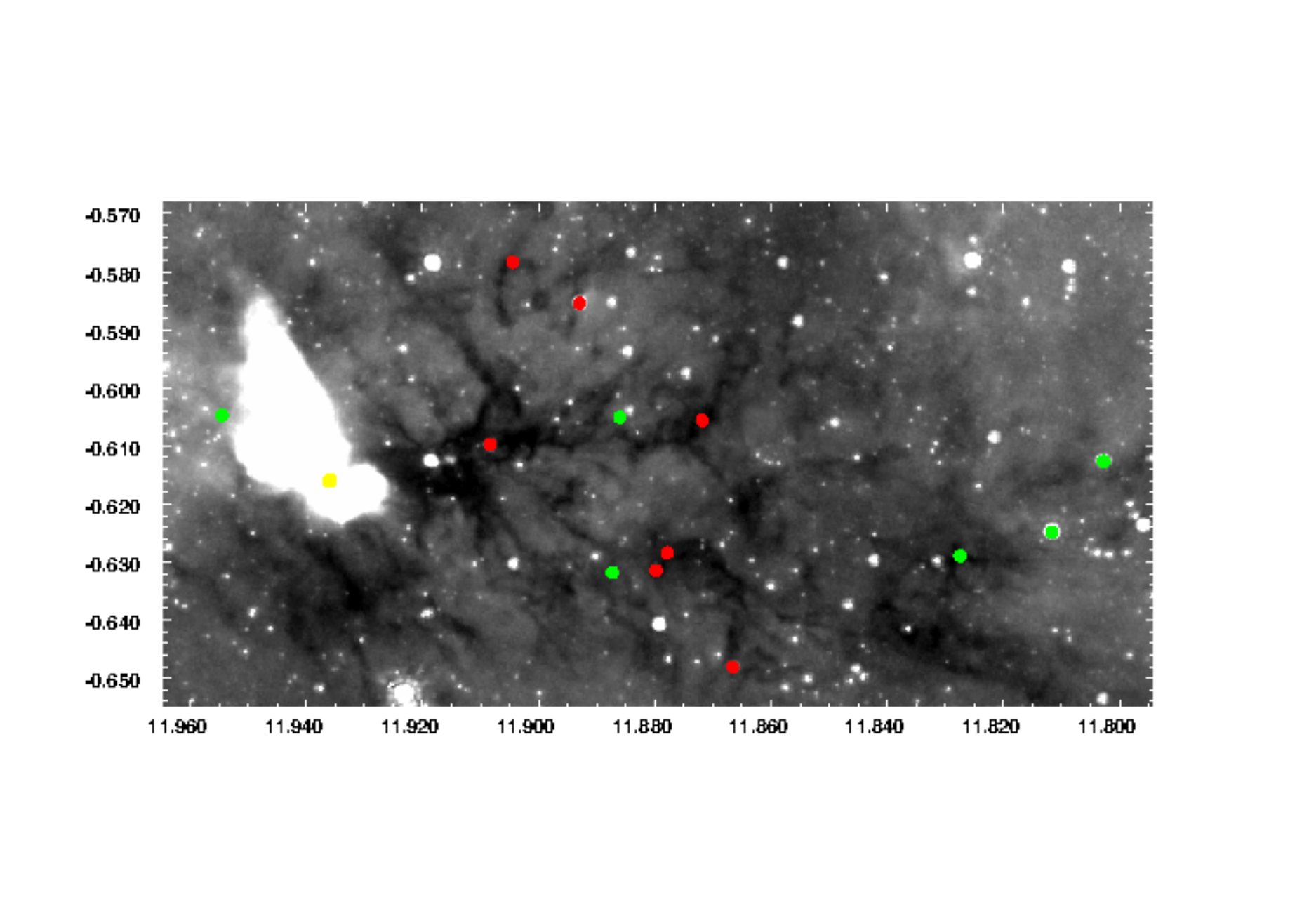}
\caption{G11.86-0.62 (Region 8)}
\end{figure*}
%% FIGURE 15 ..%%
\begin{figure*}
\centering
\includegraphics[trim=1cm 3cm 2cm 2cm, totalheight=0.3\textheight, clip=true, angle=0]{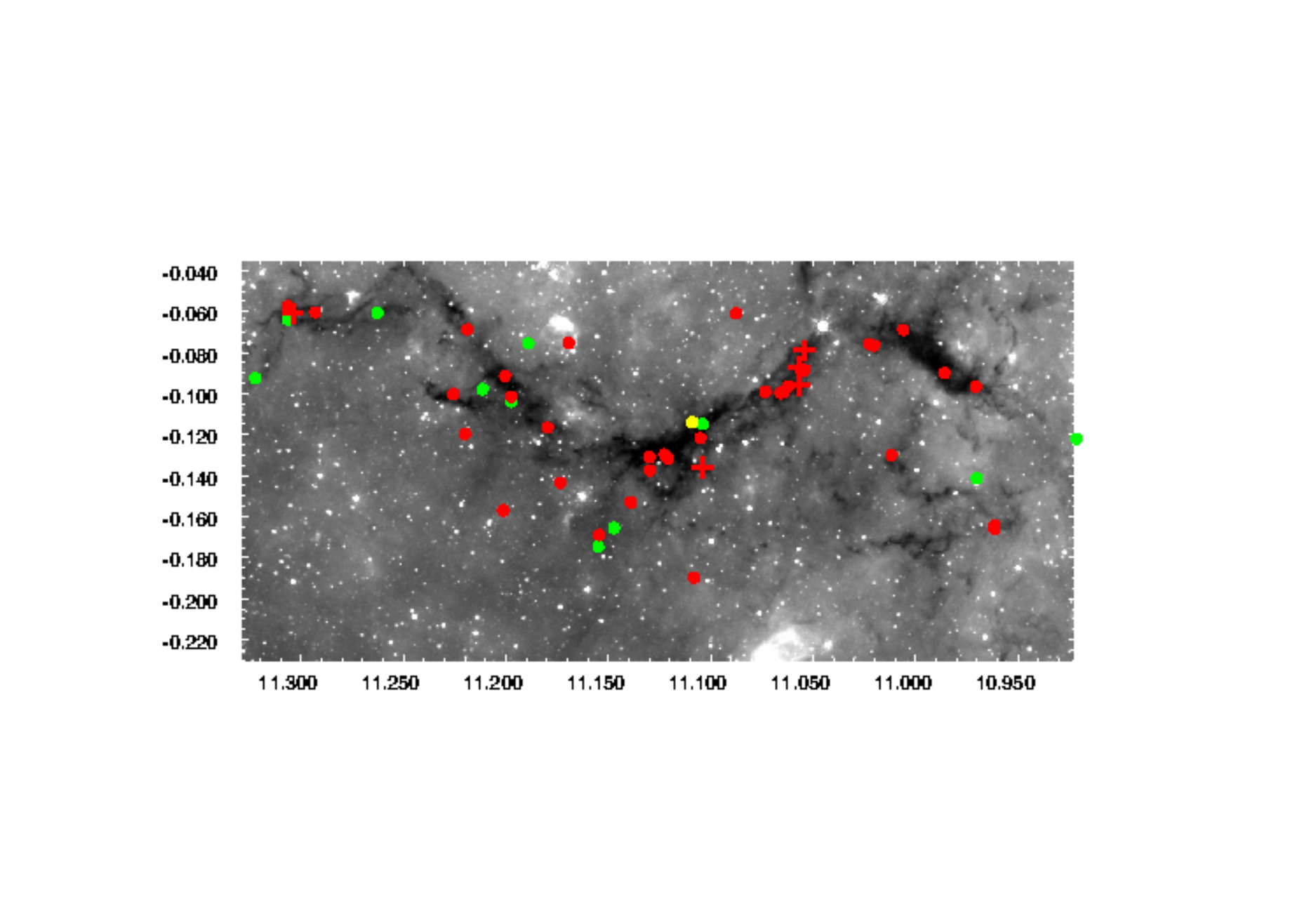}
\caption{G11.13-0.13 (Region 9)}
\end{figure*}
%% FIGURE 16 ..%%
\begin{figure*}
\centering
\includegraphics[trim=0cm 0cm 0cm 0cm, totalheight=0.4\textheight, clip=true, angle=0]{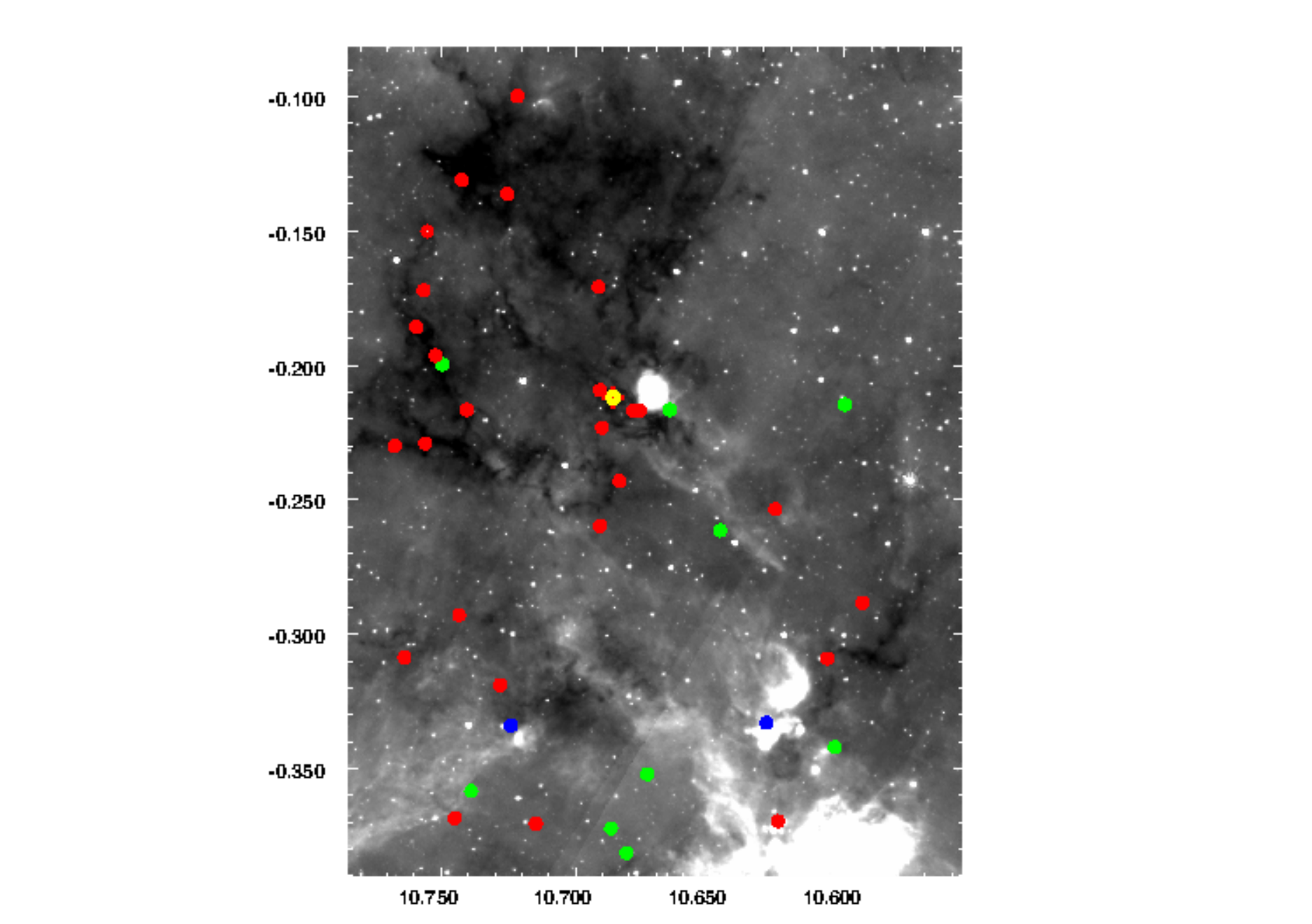}
\caption{G10.67-0.21 (Region 10)}
\end{figure*}

%% FIGURE 17 ..%%
\begin{figure*}
\centering
\includegraphics[trim=3cm 0.5cm 2cm 0cm, clip=true, totalheight=0.35\textheight, angle=0]{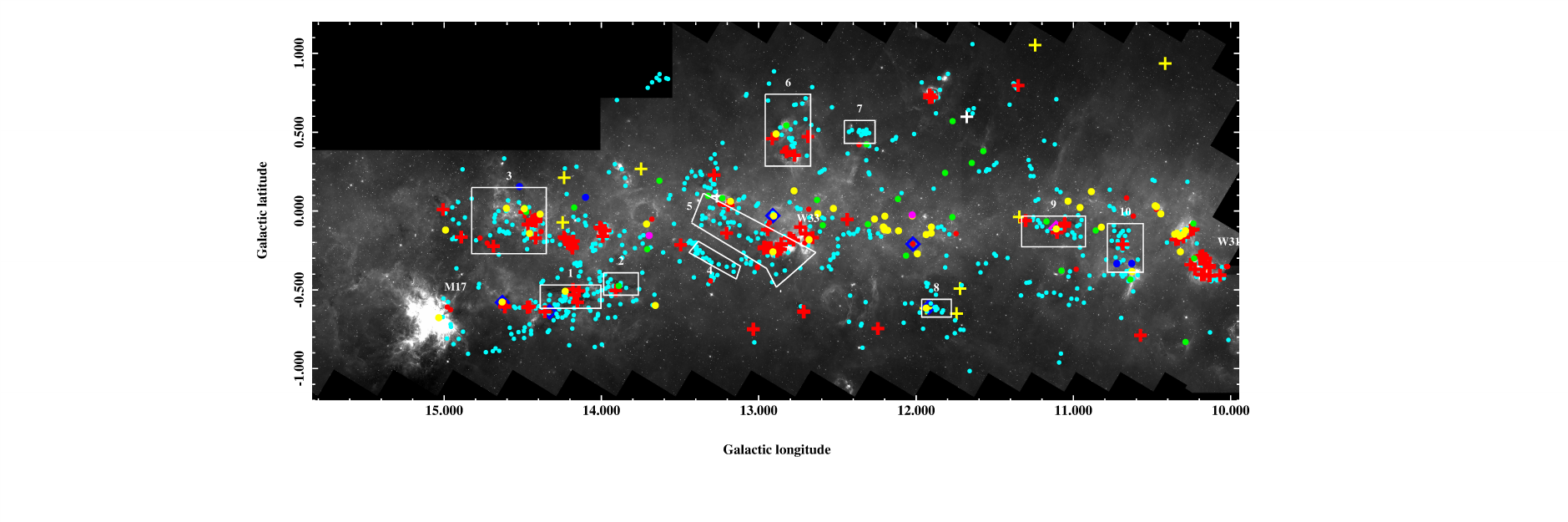}
\caption{GLIMPSE 5.8$\mu$m image of entire region with the 
location of sources which are used as 
tracers of massive star formation and distance 
informations in our study. 
Circles:- Cyan:IRDCs, Green:IR bubbles/HII regions, Yellow:Masers, 
Blue:Near 3-Kpc arm Masers, Magenta:Far 3-Kpc arm masers, 
Red crosses: Star-less clumps with near distance solutions assumed, Red circles: 
Star-less clumps with far solutions, 
Blue diamonds:'likely' MYSOs, Magenta diamonds:'possible' MYSO, White crosses:HMPOs, 
Yellow crosses:Radio sources}
\end{figure*}

%% FIGURE 18 ..%%
\begin{figure*}
\begin{center}
\subfigure{
\includegraphics[width=7cm]{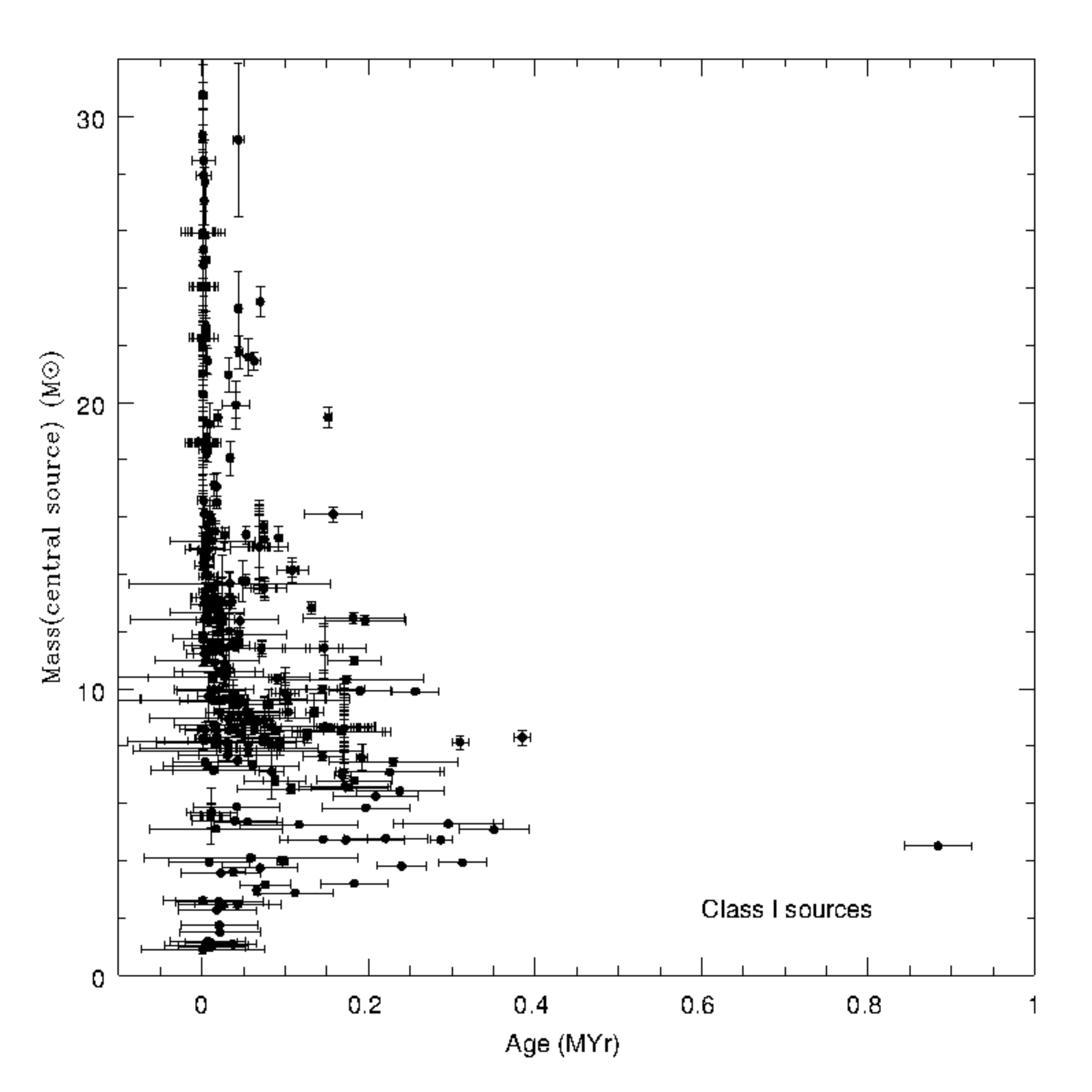}
        }
\subfigure{
\includegraphics[width=7cm]{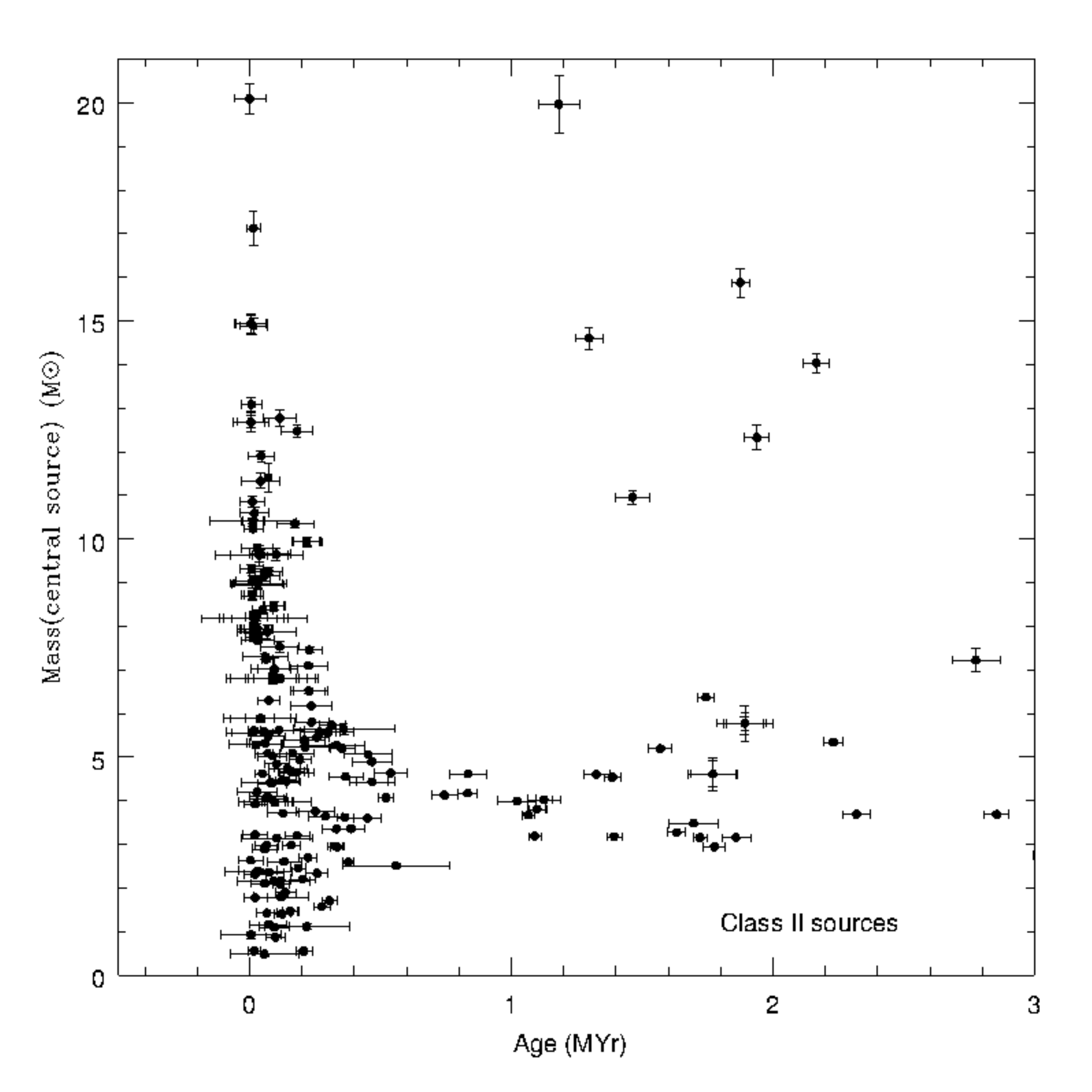}
        }\\ 
\end{center}
\caption{Mass and Age distribution of sources in the entire region. 
The central black dots correspond to the estimated mass and age of 
each of the YSOs. 
Std. deviations in estimated age and mass are shown as vectors along X axis 
and Y axis respectively.}
\end{figure*}

\end{document}